\documentclass[pdflatex]{article}
\usepackage{graphicx}
\usepackage{hyperref}
\usepackage{multirow}
\usepackage{amsmath,amssymb,amsfonts}
\usepackage{amsthm}
\usepackage{mathrsfs}
\usepackage[title]{appendix}
\usepackage{xcolor}
\usepackage{textcomp}
\usepackage{manyfoot}
\usepackage{booktabs}
\usepackage{algorithm}
\usepackage{algorithmicx}
\usepackage{algpseudocode}
\usepackage{listings}
\usepackage{makecell}
\usepackage{float}
\usepackage{placeins}
\usepackage{authblk}
\usepackage[font=footnotesize]{caption}
\usepackage[a4paper,top=1in,bottom=1in,left=0.75in,right=0.75in]{geometry}
\usepackage[backend=bibtex,style=nature,date=year]{biblatex}
\begin{document}
\renewcommand{\Affilfont}{\small}
\title{\huge{\textbf{Quantum Chemistry Driven Molecular Inverse Design with Data-free Reinforcement Learning}}}

\author[1,2]{Francesco Calcagno\thanks{\href{mailto:francesco.calcagno@unibo.it}{francesco.calcagno@unibo.it}}}
\author[3]{Luca Serfilippi}
\author[3]{Giorgio Franceschelli}
\author[1]{Marco Garavelli}
\author[3,4]{Mirco Musolesi}
\author[1,2]{Ivan Rivalta\thanks{\href{mailto:i.rivalta@unibo.it}{i.rivalta@unibo.it}}}

\affil[1]{Department of Industrial Chemistry, Alma Mater Studiorum University of Bologna, via Piero Gobetti 85, 40129 Bologna, Italy}
\affil[2]{Center for Chemical Catalysis - C3, Alma Mater Studiorum University of Bologna, via Piero Gobetti 85, 40129 Bologna, Italy}
\affil[3]{Department of Computer Science and Engineering, Alma Mater Studiorum University of Bologna, Viale del Risorgimento 2, 40136 Bologna, Italy}
\affil[4]{Department of Computer Science, University College London, 66-72 Gower Street, WC1E 6BT, London, United Kingdom}

\date{}

\maketitle

\begin{abstract}
    The inverse design of molecules has challenged chemists for decades. In the past years, machine learning and artificial intelligence have emerged as new tools to generate molecules tailoring desired properties, but with the limit of relying on models that are pretrained on large datasets. Here, we present a data-free generative model based on reinforcement learning and quantum mechanics calculations. To improve the generation, our software is based on a five-model reinforcement learning algorithm designed to mimic the syntactic rules of an original ASCII encoding based on the SMILES one, and here reported. The reinforcement learning generator is rewarded by on-the-fly quantum mechanics calculations within a computational routine addressing conformational sampling. We demonstrate that our software successfully generates new molecules with desired properties finding optimal solutions for problems with known solutions and (sub)optimal molecules for unexplored chemical (sub)spaces, jointly showing significant speed-up to a reference baseline.
\end{abstract}

\newpage
\section{Introduction}
The inverse design of new molecules \cite{freeze2019} is one of the leading challenges of chemistry this century \cite{aspuruguzik2018, mroz2022, poree2017}, aiming to generate \textit{de novo} compounds with desired properties \cite{freeze2019}. This is a fundamental paradigm shift in computational chemistry that promises a fast discovery of, e.g., new catalysts, drugs, molecular energy storage, and carbon-capturing systems.

However, the complex structure-property relationship in molecules \cite{freedomdesign} and the lack of a unifying theory to solve this problem limit its development. 
Moreover, brute-force approaches are computationally infeasible due to the exponential size of the chemical space (CS) \cite{kirkpatrick2004}. Despite significant efforts in developing physics-based methods as reported by Batista \cite{tio2_VSB, ni_VSB} and Reiher \cite{reiher_gdmc, reiher_gdmc2},  machine learning (ML)-based methods have recently emerged as powerful tools to accelerate the generation of molecules with predefined properties \cite{butler2018, sanchezlengeling2018, schwalbekoda2020, bilodeau2022, anstine2023, gow2022, jcc_rl}. 
However, ML models rely on large datasets and do not guarantee a thorough exploration of the CS, leaving a general and data-free approach to inverse design molecules an open question \cite{anstine2023}. Among different generative models, those based on reinforcement learning (RL) \cite{sutton2018} are extremely promising \cite{gow2022, anstine2023, jcc_rl}. In RL, an artificial agent learns an optimal policy to exploit a task by interacting with its environment through a trial-and-error procedure. In the inverse design of molecules, an RL agent learns how to generate molecules that maximize desired properties, hereafter called \textit{chemical reward} ($r_c$).
To the best of our knowledge, a limited number of remarkable examples of data-free generations of molecules based on RL have been reported to date. Those models adopt reward metrics based on physicochemical properties, like the drug-likeliness (QED) \cite{qed} or lipophilicity (estimated by the logarithm of the partition coefficient, logP) \cite{logP, zhou2019, thiede2022}, which are not based on quantum mechanics (QM) first-principles. QM-driven generation was recently reported for organic electronic molecular design, but using pretrained language models \cite{rldft}. Therefore, a method for QM-driven fully data-free generation of molecules is missing, representing a major gap in the field. 

In this work, we present PROTEUS, a new tool for data-free RL algorithm for molecular inverse design employing on-the-fly QM calculations of the target property.
We successfully applied PROTEUS to design chemical substituents for a molecular backbone to maximize energy difference between two geometrical isomers -- hereafter referred to as the \textit{isomerization energy}. We demonstrated that PROTEUS enables an extensive and effective exploration of highly challenging and chemically-relevant fully-characterized spaces of solutions comprising up to 2,430,845 solutions, paving the way for a new paradigm in the \textit{de novo} generation of molecules.

\section{Results and discussion}

\subsection{The P-SMILES syntax and the isomerization energy problem} \label{sec:problem}
\begin{figure}[!htbp]
  \centering
  \includegraphics[width=1.\linewidth]{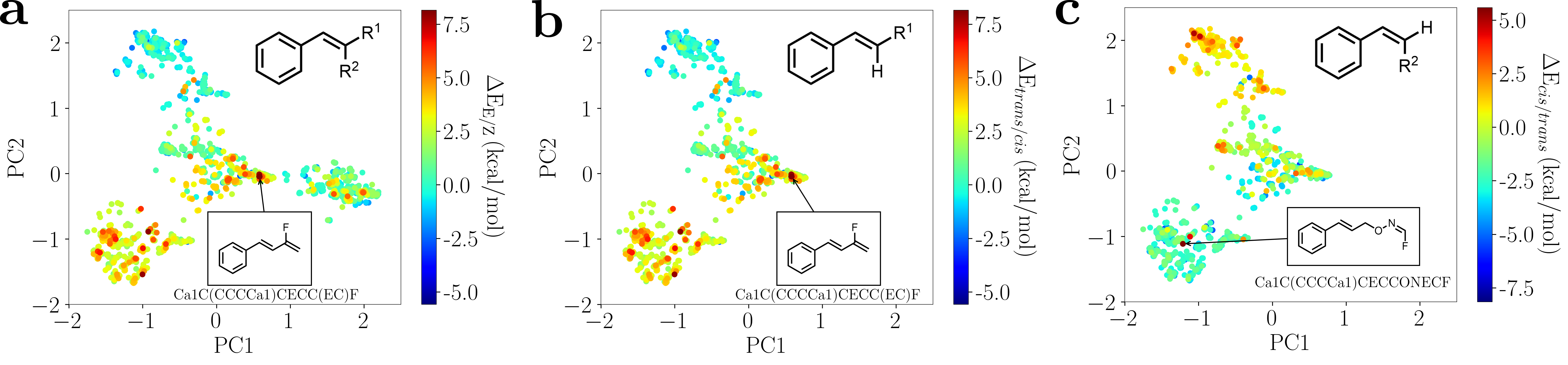}
  \caption{\textbf{Chemical space analyses and P-SMILES syntax.} Principal component analysis (PCA) of (\textbf{a}) the reference \textquoteleft E/Z dataset', (\textbf{b}) the \textit{trans}/\textit{cis} subspace (i.e. with R$^2$=H ) and (\textbf{c}) the \textit{cis}/\textit{trans} subspace (i.e. with R$^1$=H ) the. PCAs use Morgan fingerprints of Z molecules with color coding based on the energy gaps computed on DFT-optimized geometries. The structural formula and the P-SMILES strings of the molecules with the largest energy gaps, i.e. \texttt{Ca1C(CCCCa1)CECC(EC)F} (\textbf{a} and \textbf{b}) and \texttt{Ca1C(CCCCa1)CECCONECF} (\textbf{c}), are shown. 
  PCAs were done encoding molecules in bitvectors using the Morgan fingerprint scheme (radius = 5 and 4,096 bits) as implemented in the RDKit package \cite{rdkit}.}
  \label{fig:panel1}
\end{figure}

In the present work we targeted the maximization of the isomerization energy of the double C=C bond of a styrene backbone by inversely designing tailored substituents (see the inset in Fig. \ref{fig:panel1}a). The structure-property relationship for such a relatively simple molecular system -- i.e. the correlation between the structure of styrene derivatives and the corresponding isomerization energy --  is not trivial since the geometrical isomerization involves a double bond that is conjugated with an aromatic ring. 

Molecules are encoded using P-SMILES, an ASCII encoding scheme here introduced (see Sec. \ref{sec:methods}).  
P-SMILES is a SMILES-based \cite{smiles1, smiles2, smiles3} syntax that encompasses a less complex and more compact syntax, and that mitigates sources of bias during the generative RL simulations (see Sec. \ref{sec:methods}). P-SMILES simplifies the encoding by limiting to two the maximum number of tokens required to define any structural moiety, i.e. it uses either single- or double-character notation. This reduces significantly the syntactical complexity of encoding geometrical isomers and aromatic rings (see Tab. \ref{tab:syntax}), and, thus, the sources of bias in the generative RL procedure introduced by inequalities inherent to SMILES (see Sec. \ref{sec:psmiles}). Thus, as detailed in Sec. \ref{sec:methods}, PROTEUS's generative model is designed to fit the well-defined syntax properties of P-SMILES.

We approached a rigorous assessment of PROTEUS by a direct comparison between the set of molecules generated during RL experiments and the corresponding complete space of possible solutions. To keep this comparison computationally feasible, we considered complete chemical subspaces (SubCS) featuring different dimensions. The largest dataset considered, hereafter the reference \textquoteleft E/Z dataset', involves all possible sets of R$^1$ and R$^2$ substituents for the styrene's backbone resulting from the combinations of maximum 6 P-SMILES tokens. The reference dataset contains 1,628 chemically meaningful pairs of E/Z isomers out of all possible syntactic combinations, i.e. 1,948,716 pairs (see Appendix \ref{sec:secA2}).
The E/Z isomerization energy of each pair in the reference dataset was computed through a multi-step routine that accounts for QM calculations (see Sec. \ref{sec:methods}). 

Interestingly, the distribution of isomerization energies of the molecular pairs within the space of solutions highlights the complexity of the problem under investigation. Clustering of the molecules with principal component analysis (PCA) shows that the complete set of molecules can be grouped in four (Fig. \ref{fig:panel1}a) or three -- when R$^2$=H  (Fig. \ref{fig:panel1}b) or R$^1$=H (Fig. \ref{fig:panel1}c) -- main clusters with similar molecular features (see Sec. \ref{sec:clustering}). In each cluster, however, the distribution of energy gaps is heterogeneous, i.e. clusters contain both positive and negative values.

\FloatBarrier
\subsection{Inverse design of molecules with PROTEUS} \label{sec:proteus_workflow}
\begin{figure}[!htbp]
  \centering
  \includegraphics[width=1.\linewidth]{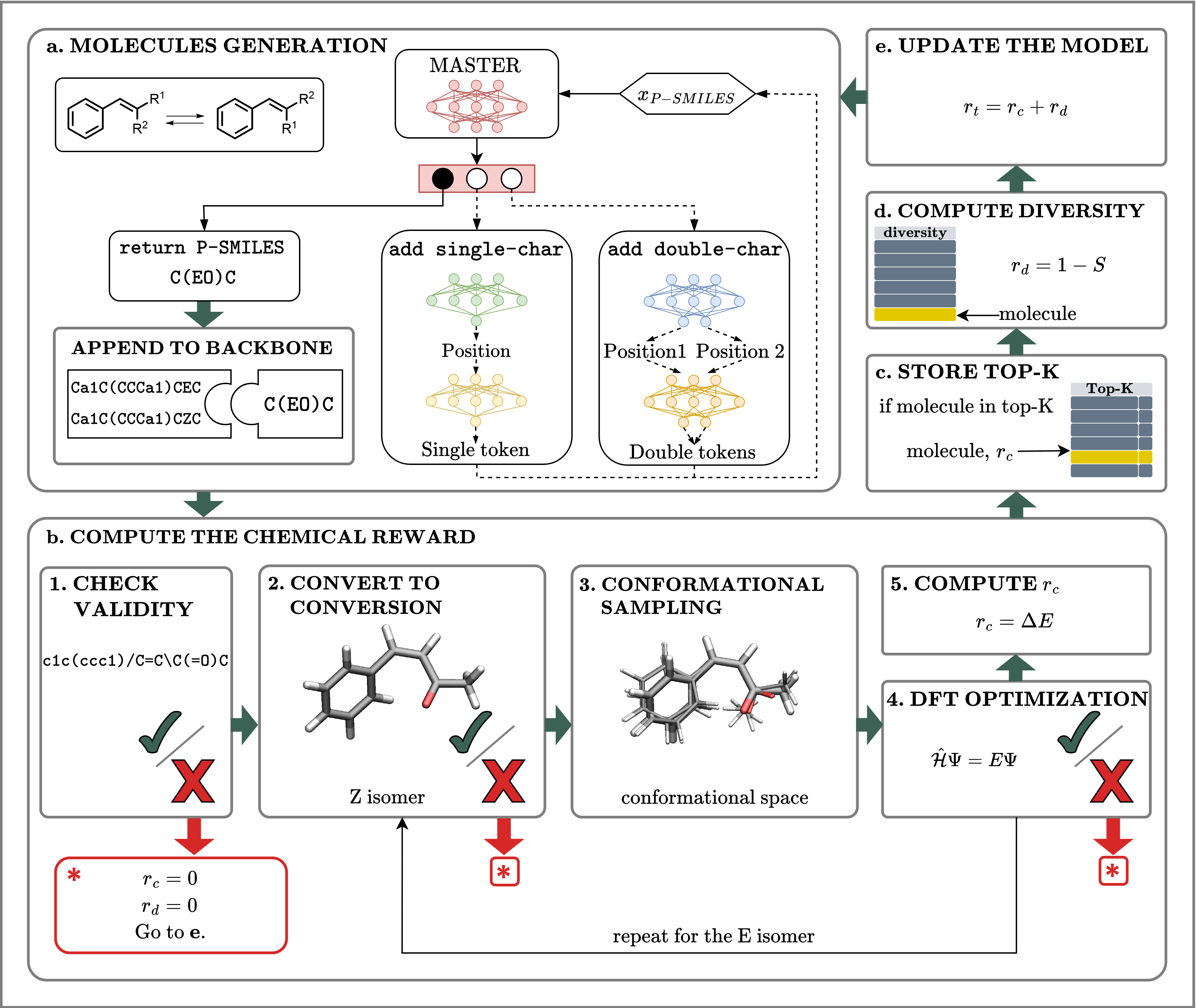}
\caption[PROTEUS]{\textbf{Data-free generation of molecules with PROTEUS.} \textbf{a}, Two substituents, i.e. R$^1$ and R$^2$ groups, for the styrene backbone (see the inset) are embedded in a P-SMILES string. This string is iteratively generated using a five-model RL agent algorithm, comprising a master decision-maker and single- and double-character predictors. The selected P-SMILES string is then appended to the styrene backbone, as shown for a simple exemplifying case, i.e. R$^1$=COCH$_3$ and R$^2$=H for the E isomer (and vice versa for the Z isomer). \textbf{b}, The molecule's (i.e. state's) chemical reward $r_c$ is computed with the following procedure. In step \textbf{1}, the P-SMILES string is first converted to a SMILES string. Then, if the SMILES has been previously generated, the $r_c$ value is not computed again, moving directly to \textbf{c}; otherwise, a syntactic validity check is performed. If the syntax is not valid, the molecule is considered invalid, and the total reward, $r_t$, is null. If the syntax is correct, the SMILES string is converted to Cartesian coordinates and it is pre-optimized at MM level. In step \textbf{2}, a geometry optimization at the  DFT-TB level is performed and if changes in the connectivity occur, the molecule is considered invalid. Otherwise, in step \textbf{3} a conformational sampling is performed using MTMD. In step \textbf{4}, the most stable conformer is optimized at the DFT level. If any structural change occurs, the P-SMILES string is considered invalid. Steps 1-4 are performed for both the E and the Z isomers and then (step \textbf{5.}) the E/Z energy gap between isomers (i.e. $r_c$) is computed. \textbf{c}, If the molecule is among the best K molecules generated so far, it is added (as marked in yellow) to the top-K memory to prioritize training toward more effective solutions. \textbf{d}, The diversity reward, $r_d$, is computed as the complementary of the Tanimoto similarity (\textit{S}). \textbf{e}, The overall reward is calculated and the PPO algorithm is used to train the five models.}
  \label{fig:proteus} 
\end{figure}

The data-free inverse design strategy of PROTEUS is depicted in Fig. \ref{fig:proteus}. PROTEUS is an RL-based model that generates molecules encoded in P-SMILES strings. It is based on the proximal policy optimization (PPO) \cite{schulman17} scheme to learn how to generate new molecules to maximize the outcome of QM calculations. RL solves generative modeling tasks formulated as Markov decision processes, i.e. defined in terms of states, actions, and rewards \cite{franceschelli2024}. The P-SMILES string encoding a molecule is the state ($s_t$), and the agent leverages a complex architecture that fits the characteristics of P-SMILES strings. Namely, since a character-based sequential generation could penalize features encoded by two characters, such as cycles and branches, our architecture is hierarchical \cite{hierarchical} for molecule generation. As illustrated in Fig. \ref{fig:proteus}a, the agent is composed of five neural network models, i.e. five NNs: \textit{i}) a master decides whether to add single-characters, double characters or to end the generation; two positional predictors decide where to place \textit{ii}) a single character or \textit{iii}) a double character in the current P-SMILES string; and two generators to effectively add a \textit{iv}) single character or \textit{v}) double character. Therefore, the action space is different for each model: the master leverages three possible actions, while both the positional predictors and the tokens generators return the numerical positions and the vocabulary tokens, respectively, depending on whether they work with single- or double-character.
The cost function, $r_t$ (see Sec. \ref{sec:methods}), rewards the generated valid molecules considering both the target chemical property (i.e. the isomerization energy), $r_c$, and a chemical diversity index, $r_d$, as follows:
\begin{equation}
    r_t = \alpha r_c\!\left(s_t\right) + \beta r_d\!\left(s_t\right), 
    \label{eq:rfitness-simple}
\end{equation}
where $r_d$ is defined as the reciprocal number of the Tanimoto similarity \cite{tanimoto} (see Sec. \ref{sec:diversity}), while $\alpha$ and $\beta$ are hyperparameters (see Sec. \ref{sec:sensitivity_analysis}). 
$r_c$ and $r_d$ cooperate in the learning process. In fact, a key ingredient of our generative model is the balance between an efficient exploration of the CS through rewarding the chemical diversity, $r_d$,  and a proper exploitation of the target chemical reward by maximizing $r_c$.

To push the exploration of unknown regions of the CS, i.e. avoiding the RL generator from being trapped in local minima, an entropy term is added to the loss function (see Sec. \ref{sec:methods}). The entropy bonus aims to include noise into the generative decision process and it avoids a deterministic choice of actions, being informed of low-explored regions of the CS. At the same time, the architecture of PROTEUS prioritizes the training towards solutions that have proven to be (sub-)optimal. Namely, PROTEUS stores the top-K P-SMILES strings generated so far, focusing the training on those solutions by doubling their weights in the actual training batch. This type of generator can, thus, focus on both under-explored and high-rewarded regions through the cooperation of various contributions: while the diversity and the entropy term push the exploration of the CS, the top-K strategy fosters the exploitation of the most promising chemical subspaces.

%\FloatBarrier
\subsection{Inverse Designing isomers}

% \FloatBarrier
\subsubsection{Inverse design of E/Z isomers} \label{sec:EZ_prob}
\begin{figure}[!htbp]
  \centering
  \includegraphics[width=0.7\linewidth]{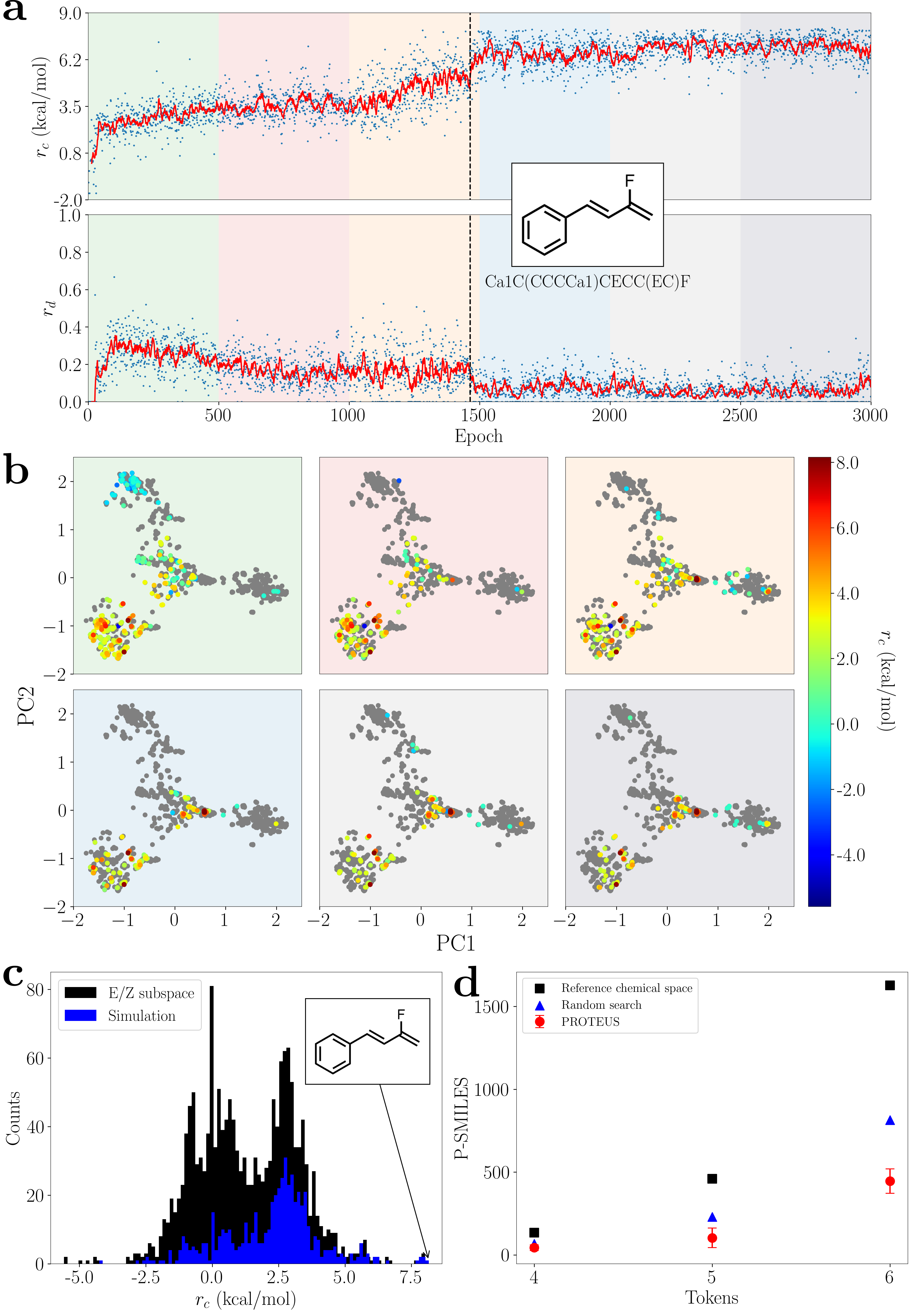}  
  \caption{\textbf{Inverse design of E/Z isomers with PROTEUS.} \textbf{a}, Time-evolution of the chemical and the diversity rewards during a representative PROTEUS simulation for the E/Z isomers within the 6-token CS. Both the mean value of each epoch (blue scatter) and the running average (solid red line) are reported. The epoch corresponding to the first generation of the best solution (with molecular formula and P-SMILES string in the inset), as ranked in the \textquoteleft E/Z dataset', is marked with a dashed line. The 3,000 epochs reported are divided into 6 windows with different background colors. \textbf{b}, For each of these 6 simulation windows, the generated molecules are displayed using the principal components defined for the full E/Z space (and reported in Fig. \ref{fig:panel1}c). The states generated in each window are labeled using the E/Z energy gap values defined by the color bar, while molecules belonging to the reference CS that are not explored are reported in grey. \textbf{c}, The E/Z energy gap distributions for isomers in the \textquoteleft E/Z dataset' (in black) and in the PROTEUS simulations (in blue) are compared. \textbf{d}, The total number of valid states in the CS (black squares) and the average number of valid generations needed to find the best solution using a random search (dark blue triangle) or PROTEUS (red circles) are compared for CS with different sizes (from 4 to 6 P-SMILES tokens). The performance of PROTEUS's outcomes reported in panel \textbf{d} are averaged over three independent simulations with three different seeds and the error bar shows the standard deviation.}
  \label{fig:panel2}

\end{figure}

Fig. \ref{fig:panel2} shows a representative PROTEUS simulation (simulation 9, Tab. \ref{tab:simulation-results}) out of three independent experiments performed for the CS with 6 tokens (simulations 7-9, Tab. \ref{tab:simulation-results}) for the maximization of the E/Z energy gap in styrene derivatives obtained by the optimization of the R$^1$ and R$^2$ substituents (see the inset in Fig. \ref{fig:proteus}a), considering R$^1$ having always higher chemical priority than R$^2$ (for the sake of simplicity). Being asked to walk in a large field of trees while looking for the \textquoteleft best fruits', thanks to its ML architecture, PROTEUS initially performs a quite broad exploration. In the first 500 epochs, PROTEUS generates molecules featuring $r_c$ values that lie in a broad distribution, i.e. with an energy gap between -4.25 and 7.95 kcal/mol (between -4.23 and 7.96 kcal/mol on average for simulations 7-9), as a direct consequence of random initialization of the policy (Fig. \ref{fig:panel2}a). Notably, the quality of this broad exploration is corroborated by an average E/Z energy gap value of $r_c$ (2.85, 2.45 and 3.52 kcal/mol in simulation 9, 8 and 7, respectively, see Tab. \ref{tab:simulation-results}) that is close to that of the whole CS (1.33 kcal/mol). The broad exploration is also witnessed by a large chemical diversity, $r_d$, value of the explored states. In fact, the running average of $r_d$ reaches its maximum value (0.35; 0.34 on average for 7-9) in these first 500 epochs (Fig. \ref{fig:panel2}a). As shown in Fig. \ref{fig:panel2}b, the valid P-SMILES strings generated during the first 500 epochs belong to all clusters composing the reference CS.

In the next 500 epochs, the exploration prioritizes regions featuring larger reward values, with the average $r_c$ value increasing to 3.56 kcal/mol, and reducing the diversity of the states (with $r_d$ averaging to 0.17), as depicted in Fig. \ref{fig:panel2}b. In the 1,000-1,500 epochs region, while keeping a quite constant diversity value in the exploration (with an average $r_d$ of 0.16), PROTEUS largely exploits the chemical reward, with a steep increase in $r_c$ that culminates with the generation of the \texttt{Ca1C(CCCCa1)CECC(EC)F} state, which is the molecule with the largest E/Z energy gap (i.e. 8.15 kcal/mol) in the full SubCS of solutions. After finding the very \textquoteleft best fruit' (1,500-3,000 epochs), PROTEUS mainly exploits the high $r_c$ values with a concomitant drop of the chemical diversity, i.e. it focuses on the best fruits in the best trees. In fact, the average $r_c$ ranges between 6.59 and 7.03 kcal/mol in the 1,500-3,000 epochs, while the average $r_d$ lowers down below 0.07 (see Fig. \ref{fig:panel2}b). The opposite trend of $r_c$ and $r_d$ is to be ascribed to the cooperation between the exploration and exploitation during the learning process. During the exploration phase, when $r_c$ values are low, the impact of $r_d$ on the final value of $r_t$ (Equation \ref{eq:rfitness-simple}) is not negligible. Instead, when PROTEUS exploits the chemical reward $r_c$, the weight of $r_c$ becomes much larger than $r_d$, so limiting the impact of the exploration. 

A similar behavior in the exploitation of $r_c$ was observed in the other replicas of the same simulation, but showing different time scales (see Supporting Information). 

Comparing the distribution of energy gaps in the complete \textquoteleft E/Z dataset' with those of the states explored by PROTEUS during the 3,000 epochs of simulation 9, as depicted in Fig. \ref{fig:panel2}c, provides further insights into the learning process. PROTEUS clearly overall prioritizes the generation of valid states with high $r_c$ values, exploring primarily regions with chemical rewards larger than ca. 2 kcal/mol, and providing a great computational speed up in the search for the best pair of E/Z isomers. To evaluate the characteristic computational advantage of PROTEUS we compared it with a random search approach for SubCSs composed of 4-, 5-, and 6-token P-SMILES molecules. For each SubCS, three independent simulations were carried out (see Tab. \ref{tab:simulation-results}). As reported in Fig. \ref{fig:panel2}d, to find the best solution PROTEUS generates a number of unique valid P-SMILES strings that is very close to that of a random search (determined as half of the total solutions) only when the size of the CS of valid solutions is small. 
The 4-token SubCS is composed of 134 unique valid states out of 3,770 combinations. With the random search approach, 67 random valid generations are required on average before sampling the best solution. Similarly, PROTEUS required on average 43$\pm$17 unique valid P-SMILES (Tab. \ref{tab:simulation-results}). In the cases of the 5- and 6-token spaces, instead, PROTEUS successfully generates the best molecule after generating on average 103$\pm$59 and 445$\pm$75 unique and valid samples, respectively (Fig. \ref{fig:panel2}d). Since the random search approach requires 229 and 814 iterations, respectively, PROTEUS's results should be considered outstanding, since it drastically reduces the number of expensive QM property evaluations before finding the best solution. 

% \FloatBarrier
\subsubsection{Inverse design of \textit{trans}/\textit{cis} and \textit{cis}/\textit{trans} isomers} \label{sec:CT_prob}
The second inverse design problem we tackled with PROTEUS is the design of a tailored substituent R$^1$ that maximizes the \textit{trans}/\textit{cis} energy gap (i.e. the stabilization of the \textit{trans} isomer) for the styrene derivatives (Fig. \ref{fig:panel1}b). The \textit{trans}/\textit{cis} problem, while being chemically simpler than the E/Z one, is more complicated from the point of view of the learning process since the new constraint does not affect the total number of combinations of P-SMILES tokens, which is the same as in the \textquoteleft E/Z dataset', whereas the valid pairs of isomers decrease from 1,628 to 1,246 (hereafter referred to the \textquoteleft \textit{trans}/\textit{cis} dataset'), reducing the density of valid states.

Fig. \ref{fig:panel3} in Appendix \ref{secA1} shows the outcome of PROTEUS simulation for the maximization of the \textit{trans}/\textit{cis} energy gap in the 6-token space. This problem features the same best solution as the E/Z one, i.e. \texttt{Ca1C(CCCCa1)CECC(EC)F}, and similar trends to what was already discussed for the E/Z simulations: after a broad exploration of the CS, PROTEUS successfully focuses on maximizing $r_c$ up to generating the best molecule.

To further assess the capabilities of PROTEUS in balancing exploration and exploitation, we tested the inverse design routine for the reverse (energetic) inverse design problem, i.e. the maximization of the \textit{cis}/\textit{trans} energy gap. This task is significantly challenging in the context of the 6-tokens CS because the best solution has a chemical structure that is similar to molecules with much worse chemical reward, i.e. the \textquoteleft best fruit' is in the \textquoteleft worst tree' (Fig. \ref{fig:panel1}c). Despite this, PROTEUS solves the problem, confirming the virtuous balance between exploration and exploitation (see Supporting Information).

\FloatBarrier
\subsubsection{Exploration beyond a reference chemical space} \label{sec:CT_7T}  
\begin{figure}[!htbp] 
\centering
    \includegraphics[width=1.\linewidth]{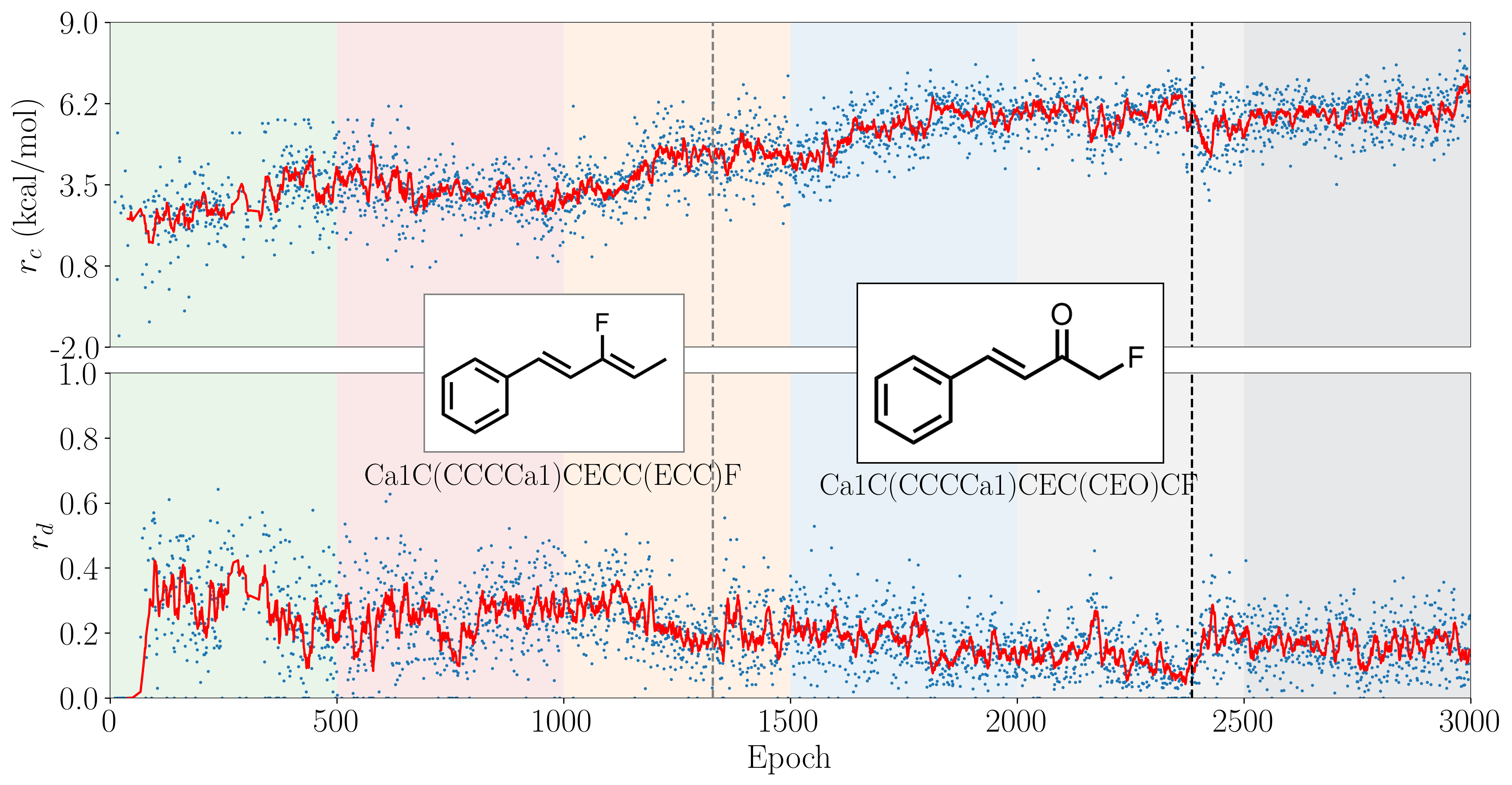}
\caption{\textbf{Exploration of a large \textit{trans}/\textit{cis} chemical space with PROTEUS.} Time-evolution of the chemical and diversity rewards during a PROTEUS simulation for the \textit{trans}/\textit{cis} isomers within the CS of 7 P-SMILES tokens. Both the mean value of each epoch (blue scatter) and the running average (solid red line) are reported. The epochs corresponding to the generations of the first solution with a \textit{trans}/\textit{cis} energy gap larger than that of the best solution found in the 6-token CS (grey dashed line) and the best solution found along the 3,000 epochs simulation (black dashed line) are highlighted, with the corresponding molecular structures and P-SMILES strings in the insets. The 3,000 epochs reported are divided into 6 windows with different background colors.}
\label{fig:sim_CT7}
\end{figure}

Given the capabilities of PROTEUS in solving the molecular inverse design problem for different cases, as shown above for CSs with known solutions, we pushed our tool to tackle an inverse design problem for which the characterization of the full reference space would require a very large computational cost with standard academic facilities (see Supporting Information). In particular, we carried out the \textit{trans}/\textit{cis} inverse design simulation with a maximum number of tokens for the generated P-SMILES states increased from 6 to 7.
Despite a larger space, for sure PROTEUS should generate sub-optimal solutions featuring larger (or at least equal) $r_c$ than in \texttt{Ca1C(CCCCa1)CECC(EC)F}, which is the global solution of the 6-token inverse design problem.
Fig. \ref{fig:sim_CT7} shows the PROTEUS simulation for the 7-token CS. As for the previous simulations, PROTEUS initially performs a broad exploration of the space of solutions but, as expected, this exploration period gets longer as the CS increases. In fact, the average $r_d$ in the first 1,000 epochs is constant at ca. 0.25, while the average $r_c$ value is 3.17 kcal/mol. In the subsequent 500 epochs, the average $r_c$ increases to 4.19 kcal/mol, and the \texttt{Ca1C(CCCCa1)CECC(ECC)F} state is generated. This molecule is quite similar to the best solution of the 6-token CS, which is generated only towards the end of the current simulation, differing just for a methyl group. These two states feature similar \textit{trans}/\textit{cis} energy gaps, whereas the 7-token solution has a slightly higher value, i.e. 8.21 kcal/mol. This achievement is remarkable since PROTEUS generates a highly-rewarded 7-token candidates without having fully explored the 6-token subspace, and requires almost the same number of epochs as for the 6-token problems (\textless 1,500). 

In the next 1,500 epochs, PROTEUS maximizes the \textit{trans}/\textit{cis} energy gap. This is witnessed by an average $r_c$ value of 5.78 kcal/mol and by the fact that the average $r_d$ decreases below 0.16 (see Fig. \ref{fig:sim_CT7}). At this stage, the \texttt{Ca1C(CCCCa1)CECC(EO)CF} state is generated, which is the best solution for the whole simulation featuring a \textit{trans}/\textit{cis} energy gap of 9.55 kcal/mol, i.e. 1.40 kcal/mol larger than \texttt{Ca1C(CCCCa1)CECC(EC)F}. 
The fast generation of such a sub-optimal solution within a challenging SubCS confirms that PROTEUS can successfully inversely design molecules also in large reference spaces, by targeting the energy gap value computed at the QM level.

\FloatBarrier
\section{Conclusions}
We proposed an ML tool for data-free \textit{de novo} generation of molecules that combines on-the-fly QM calculations and a new ASCII encoding. 

We targeted the molecular inverse design problem of maximizing the electronic energy gap between geometrical isomers using as reference the backbone the styrene molecule. This one can isomerize along a double (C=C) bond that is conjugated with an aromatic ring, leading to intricate combinations of steric and electronic effects that determine the isomerization energies. 
PROTEUS successfully disclose the best solution for a large CS that was previously fully characterized, and outperformed the random search method. The outcome demonstrated that our data-free RL technique applied to molecular inverse design problems can be successful if a good balance between exploration and exploitation is achieved during the learning process. We stress-tested PROTEUS, indeed, to solve the inverse design problem for CSs featuring \textit{i}) a smaller percentage of valid states with respect to the reference CS or \textit{ii}) a best solution with a chemical structure similar to molecules representing the worst solutions. The former problem is associated with the maximization of the \textit{trans}/\textit{cis} energy gap, which represents a simpler chemical problem than the E/Z one but, given its less dense space of valid states, leaves PROTEUS with fewer chances to learn the syntactic rules of P-SMILES in absence of a pretraining. The latter problem is, instead, the reverse energetic problem for the same CS of the former, i.e. the maximization of the \textit{cis}/\textit{trans} energy gap, which requires a virtuous balance between exploration of the space of solutions and exploitation of the task. By solving both inverse design problems brilliantly, for which the exact solution is known, PROTEUS proved to be robust and to feature enough flexibility to tackle the exploration of different CSs, as it can effectively exploit a chemical reward in multiple search directions within a CS. 

Considering the computational efforts required by brute-force and high-throughput approaches, PROTEUS provides significant computational savings that allow the exploration of large and complex CSs. We further prove it by solving the 7-token \textit{trans}/\textit{cis} problem, for which the full characterization of the CS is computationally quite demanding. Despite the best solution for this CS being unknown, PROTEUS quickly generated a 7-token solution that is better than the best solution of the 6 tokens CS, finding progressively better solutions of the inverse design tasks at an affordable computational cost for standard computational chemistry laboratories.

In conclusion, the proposed combined features of data-free RL technique and on-the-fly computations of chemical rewards at the QM level make PROTEUS a very valuable tool for solving molecular inverse design problems. PROTEUS architecture involves models that are general and can be easily adapted to exploit more complicated inverse design tasks, setting up quantum chemistry-based inverse design of molecules using data-free approaches.

\FloatBarrier
\section{Methods} \label{sec:methods}

\subsection{PROTEUS}
The RL model shown in the present work consists of five ML models. Each model implements a policy by means of a transformer architecture \cite{vaswani2017}. The models are organized as follows: the master, with policy $\pi_{M}\!\left(s_t\right)$, receives as input the P-SMILES string, $s_t$, produced so far and decides among three actions: \textit{i}) to add a single-character token to $s_t$, \textit{ii}) add double-character token to $s_t$, or \textit{iii}) return $s_t$, i.e. ending the generation. 
If the first action is chosen, $s_t$ is fed into the single-character position predictor, $\pi_{P^S}!\left(s_t\right)$. This predictor outputs a probability vector, from which the position for placing a single-character token is sampled. Therefore, the single-character generator, $\pi_{G^S}\!\left(s_t\right)$, returns a vector of probabilities from which the single-character token to be placed in the position chosen by the previous NN is sampled, and $s_{t+1}$ is obtained by modifying $s_t$ accordingly.
If $\pi_M(s_t)$ selects the second action, $s_t$ is passed to the double-character position predictor, $\pi_{P^D}\!\left(s_t\right)$, which returns two vectors of probabilities, one for each position to be chosen. Thus, the two positions are sampled to ensure always syntactically valid two-character tokens. At this stage, the double-character generator, $\pi_{G^D}\!\left(s_t\right)$, samples a two-character token, and $s_{t+1}$ is obtained.
Finally, if the last action is selected, the generation is considered as concluded. 

The architecture of PROTEUS overcomes the sequential (i.e. left-to-right) construction strategy of molecules, since it relies on a modification policy similar to the masked language modeling \cite{devlin18}. In fact, given an intermediate P-SMILES string, i.e. $s_t$, each action can modify $s_t$ by adding a single- or a double-character token in any position. This means, for example, that a chain of carbons \texttt{CCCCCC} could be easily branched (e.g. \texttt{CCC(CC)C}) after its construction by a single action, allowing PROTEUS to define complicated structures with single actions.

At the beginning of each PROTEUS simulation, the parameters of each NN are initialized randomly using the default initializer based on the Glorot uniform distribution. The complete pseudocode for the generative loop is provided in Algorithm \ref{alg:gen}.

Each policy is trained using PPO with prioritized experience replay \cite{schaul2016prioritized}. The prioritization scheme implemented in PROTEUS simply doubles the sampling probability for top-K trajectories. The overall loss for each policy is defined as follows:
\begin{equation} \label{eq:ppo}
    L\!\left(\theta\right) = \hat{\mathop{\mathbb{E}}}_t \!\left[ L^{CLIP}_t\!\left(\theta\right) - L^{VF}_t\!\left(\theta\right) + c_e S\!\left[\pi_{\theta}\right]\!\left(s_t\right) \right],
\end{equation}
\noindent 
with
\begin{equation} \label{eq:clip}
    L^{CLIP}_t\!\left(\theta\right) = \hat{\mathop{\mathbb{E}}}_t \!\left[ \min\!\left(\frac{\pi_{\theta}\!\left(a_t | s_t\right)}{\pi_{\theta_{old}}\!\left(a_t | s_t\right)} \hat{A}_t, clip\!\left(\frac{\pi_{\theta}\!\left(a_t | s_t\right)}{\pi_{\theta_{old}}\!\left(a_t | s_t\right)}, 1 - \epsilon, 1 + \epsilon\right) \hat{A}_t\right)  \right]
\end{equation}
\noindent 
and
\begin{equation} \label{eq:value}
    L^{VF}_t\!\left(\theta\right) = \hat{\mathop{\mathbb{E}}}_t \!\left[ \left( V_{\theta}\!\left(s_t\right) - V^{target}_t \right)^2 \right].
\end{equation}
\noindent 
$L^{CLIP}_t$ is the clipped surrogate objective that modifies the policy towards the maximization of the reward while preventing too large changes, and $V_{\theta}$ is the value function used to estimate the value of the current state. $V_\theta$ is computed with another transformer-based NN model identical to the policies described above. 
The advantage term $A_t$ is estimated using the one-step temporal difference error \cite{degris2012offpolicy} as follows:
\begin{equation} \label{eq:adv}
    A_t = r_t + \gamma V_{\theta}\!\left(s_{t+1}\right) - V_{\theta}\!\left(s_{t}\right),
\end{equation}
\noindent 
and the value function is trained to approximate
\begin{equation} \label{eq:return}
    V^{target}_t = A_t + V_{\theta}\!\left(s_t\right).
\end{equation}
\noindent 
$S\!\left[\pi_{\theta}\right]$ is an entropy bonus that prevents the policy from collapsing over \textit{deterministic} solution, and $c_e$ is a hyperparameter that weights the entropy term. 
Equation \ref{eq:entropy} was used to calculate the entropy of the policy, according to information theory:
\begin{equation}\label{eq:entropy}
 S(X) := - \sum \limits_{x \in X} p\left(x\right)\log_b p\left(x)\right.
\end{equation}
\noindent 
where $X$ is our policy vector, $p(x)$ is the probability of selecting action $x$, and $b$ the number of possible actions. The entropy value $S(X)$ ranges between 0 and 1. Indeed, when it is maximized, the entropy value of the policy becomes 1, meaning that the policy is completely random, that is
\begin{equation}
        p(x) = \frac{1}{b}\quad \mathrm{for } \quad \forall x \in X 
\end{equation}
\noindent
On the contrary, when $S(X) = 0$, the policy is deterministic and only one action can be selected. In plain words, the entropy is a measure of the exploration capability of our agent in a particular state: the higher the entropy, the higher the exploration. Similarly, according to the information theory, high entropy states correspond to low information content value. This underlies that the penalty to pay for a satisfactory exploration of the action space is to lower the confidence of the information content value in the policy. 

All the models share the total reward, $r_t$, which is defined as follows:

\begin{equation} 
\label{eq:fitness}
  r_t\!\left(s_t\right) =
    \begin{cases}
      -1 & \text{if } t = T \text{ and } T = 0 \text{ or }  T > L\\ 
      0 & \text{if $t < T$ or $s_{t=T}$ is not valid}\\
      \alpha r_c\!\left(s_t\right) + \beta r_d\!\left(s_t\right) & \text{if } t = T \text{ and }s_t \text{ is valid}
    \end{cases}       
\end{equation}

\noindent 
where $t$ is a generic step of an episode, $T$ is the last one, $L$ is the maximum sequence length, $r_c$ is the targeted chemical property and $r_d$ is a measure of the diversity between the given molecule and $n$ molecules generated before (see Sec. \ref{sec:diversity}), while $\alpha$ and $\beta$ are hyperparameters which weight each term. Both $r_c$ and $r_d$ are normalized with a discount-based scaling scheme \cite{engstrom20}. 
The complete training algorithm is reported in Appendix \ref{sec:secA2}.

\subsection{Energy Gap Calculation} \label{sec:fitness} 
$r_c\!\left(s_t\right)$ is a function (or a routine) to compute the desired chemical property of $s_t$, which depends on the chemical structure encoded in the state, $s_t$. This task is exploited by external software for QM calculations. In the present work, we focused on the energy gap between geometrical isomers of the same molecule.
Once $t = T$, the routine to compute the $r_c$ is switched on. It comprises different steps to check the validity of $s_t$, based on physical-chemical criteria and QM calculations:
\begin{enumerate}
    \item the generated P-SMILES string is converted to the corresponding SMILES string. If the conversion fails due to the detection of inconsistency in syntax, $s_t$ is considered invalid.
    \item If the SMILES string contains either oxygen-oxygen or nitrogen-nitrogen bonds in linear chain systems, it is considered invalid.
    \item The compliance of the basic chemical rules in the SMILES string is verified, e.g. that the number of bonds for each atom does not exceed the theoretical limit. If any rule is broken, the SMILES string is considered invalid. This validity check is done using the RDKit software package \cite{rdkit}. 
    \item The SMILES string is converted to Cartesian coordinates and it is roughly optimized at the molecular mechanics (MM) level using the MMFF94 force field (FF) \cite{mmff94} as implemented in the Pybel module \cite{pybel} in the OpenBabel python library \cite{obabel}.
    \item If the molecule is formally not closed-shell neutral, the SMILES string is considered invalid. Molecular total charge $Q$ is computed as
\begin{equation}\label{eq:charge}
        Q = \frac{\sum_i n_i Z_i}{2}
    \end{equation}
    \noindent 
    with $n$ the number of atoms of element $i$ with atomic number $Z$.
    \item A second geometry optimization of the structure is done at the DFT-TB level using the xTB software package \cite{xtb}. All optimizations have been done using the GFN2-xTB Hamiltonian \cite{gfn_1, gfn_2}.
    \item A geometry check is done on top of the optimized structure to verify that no change in the connectivity occurred during the optimization. The molecule, before and after the optimization, is converted to a graph structure, where atoms are nodes and bonds of any order are single edges. Then, the isomorphism between the graphs is verified. If the two graphs are not isomorphic, the molecule is considered invalid. It is important to highlight that a direct comparison between either SMILES or InchiKey \cite{inchikey} strings is not a valid choice at this stage since atom typing sometimes changes after the DFT-TB optimization even if the connectivity does not vary. The manipulation of graphs was done using the NetworkX python library \cite{networkx}. 
    \item The conformational analysis is done with the CREST software \cite{crest} on top of the optimized geometry. CREST relies on an automatic MTMD scheme to sample and select conformers of a given molecule. The simulation time of the MTMD is set three times longer than the default value to improve the exploration of the conformational space. Among the final ensemble of conformers, the conformer with the lowest energy is selected for the next step. 
    \item The desired molecular property is computed at the selected level of theory. In the present work, we compute the ground state electronic energy as a single point with DFT-TB or DFT level, or after optimizing the geometry of the conformer selected by CREST with DFT.
    
    All DFT calculations were carried out with the Gaussian16 software package \cite{g16} and using the exchange-correlation B3LYP functional \cite{b3lyp1, b3lyp2, b3lyp3, b3lyp4, b3lyp5} in pair with the 6-31G(d,p) basis set for all the elements \cite{631gd}. 
    \item The InchKey strings of the generated P-SMILES and of the geometry from the previous step are compared to verify that no changes occurred in connectivity, bond orders, or isomerism during the whole routine. Contrary to step 7, working with InchiKey strings is the best choice at this step to ensure an exact correspondence between the generated P-SMILES string, i.e. the state $s_t$, and its reward value.
    \item In the present work, we evaluate the isomerization energy, thus, all the steps are repeated for both the E (or \textit{trans}) and the Z (or \textit{cis}) isomers. Then, the isomerization energy is computed as the difference between the electronic energies of the two isomers as follows:
    \begin{align} 
        \Delta E_{\text{E/Z}} &= E_{\text{Z}} - E_{\text{E}} \label{eq:rfitness_EZ}
        \\
        \Delta E_{cis\text{/}trans} &= E_{\text{trans}} - E_{\text{cis}} \label{eq:rfitness_CT}
        \\
        \Delta E_{trans\text{/}cis} &= E_{\text{cis}} - E_{\text{trans}} \label{eq:rfitness_TC} 
  \end{align}
\end{enumerate}

\subsection{Diversity} \label{sec:diversity}
The diversity reward, $r_d\!\left(s_t\right)$, of a given valid molecule is calculated as the reciprocal number of the Tanimoto similarity $S\!\left(\cdot,\cdot\right)$ \cite{tanimoto} between the current state and the most similar molecule over the last $n$ valid generated molecules,
\begin{equation} \label{eq:diversity}
    r_d\!\left(s_t\right) = 1 - \underset{\forall b \in B_n}{\max}S\!\left(a\!\left(s_t\right), b\right)\\,
\end{equation}
\noindent 
with $n$ a hyperparameter and
\begin{equation} \label{eq:tanimoto}
    S\!\left(a,b\right) = \frac{\lvert a \cap b \rvert}{\lvert a \cup b \rvert}.
\end{equation}
\noindent 
$a$ and $b$ are the fingerprint arrays to which each molecule (i.e. SMILES string) is embedded based on given features \cite{tanimoto}. Fingerprints were computed using the MACCS fingerprint as implemented in the Pybel python library \cite{pybel}.

\subsection{P-SMILES}
P-SMILES streamlines SMILES  syntax \cite{smiles1, smiles2, smiles3} to simplify the syntactic complexity to define isomeric isomers and aliphatic/aromatic rings. Namely, the two- or three-character notations of SMILES of E and Z isomers is substituted with a one-character one, i.e. introducing the \texttt{E} and \texttt{Z} tokens. Similarly, aromatic rings are defined by couples of token \texttt{an} (with \texttt{n} $\in \mathbb{Z}^+$), instead of the conventional representation involving paired numbers and juxtaposed double bonds. P-SMILES, thus, reduces the number of symbols used in SMILES, while retaining its encoding capabilities (see Appendix \ref{sec:secA2} for further details). 

\subsection{The \textquoteleft E/Z dataset'} \label{sec:EZdb} 
The \textquoteleft E/Z dataset' is a complete subspace of valid molecules comprising all the possible E/Z isomers of the styrene backbone, whose substituents are encoded as P-SMILES strings with no more than 6 tokens. The set of tokens we used is the following: [\texttt{E}, \texttt{Z}, \texttt{a1}, \texttt{1}, \texttt{\#}, \texttt{(}, \texttt{)}, \texttt{C}, \texttt{N}, \texttt{O}, \texttt{F}]. 

\section*{Code availability}
The code of PROTEUS is available from the corresponding authors under request.

\section*{Author contribution}
F.C. and I.R. conceived the idea of PROTEUS. F.C. and I.R. designed the quantum chemistry routines implemented in PROTEUS. L.S., G.F. and M.M. designed the reinforcement learning framework at the basis of PROTEUS. F.C., L.S., G.F., M.M., and I.R. contributed to the design of the software and discussed results. F.C. and L.S. wrote the code, performed the experiments, and analyzed data. M.G., M.M., and I.R. provided financial support. F.C., L.S., and G.F. wrote the manuscript. M.M. and I.R. reviewed the manuscript.

\begin{appendices}

\FloatBarrier
\section{Supporting Data}\label{secA1}

\begin{table}[!htbp] 
\centering
\caption{Comparison between the SMILES and P-SMILES notations to encode geometrical isomers of azobenzene, used here as a general exemplifying case.}
\label{tab:syntax}
\begin{tabular}{@{}c|c|c@{}}
\toprule
\textbf{Molecules} & \textbf{SMILES} & \textbf{P-SMILES} \\ [0.5ex]
\midrule
\raisebox{-.5\height}{\includegraphics[width=0.3\linewidth]{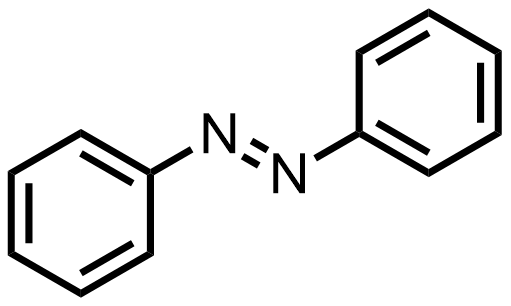}} & 
\makecell[c]{\texttt{C1=CC=CC=C1/N=N/C1=CC=CC=C1}\\
             \texttt{C1=CC=CC=C1\textbackslash N=N\textbackslash C1=CC=CC=C1}\\
             \texttt{C1=CC=CC=C1N=NC1=CC=CC=C1}\\
             \texttt{c1ccccc1/N=N/c1ccccc1}\\
             \texttt{c1ccccc1\textbackslash N=N\textbackslash c1ccccc1}\\
             \texttt{c1ccccc1N=Nc1ccccc1}} & 
\texttt{Ca1CCCCCa1NENCa1CCCCCa1} \\
\midrule
\raisebox{-.5\height}{\includegraphics[width=0.23\linewidth]{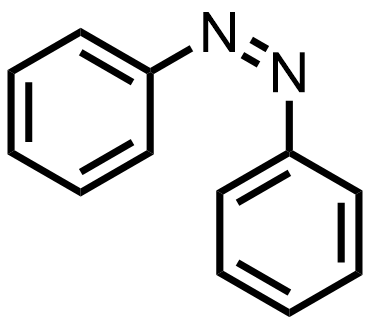}} & 
\makecell[c]{\texttt{C1=CC=CC=C1/N=N C1=CC=CC=C1}\\
             \texttt{C1=CC=CC=C1\textbackslash N=N/C1=CC=CC=C1}\\
             \texttt{c1ccccc1/N=N\textbackslash c1ccccc1}\\
             \texttt{c1ccccc1\textbackslash N=N/c1ccccc1}} & 
\texttt{Ca1CCCCCa1NZNCa1CCCCCa1} \\
\bottomrule
\end{tabular}
\end{table}
\begin{table}[!htbp] 
\caption{\textbf{PROTEUS simulations.} Results of independent RL simulations replicas. The number of epochs played before generating the best state as ranked in the reference \textquoteleft E/Z dataset', the number of unique states ($u$) before the best generation, and its average over the replicas (${\bar{u} \pm \sigma}$), with $\sigma$ the standard deviation, are reported for the 4-, 5-, and 6-token E/Z PROTEUS simulations. The first number of the simulation label indicates the number of P-SMILES tokens involved.}
\label{tab:simulation-results}
\centering
\begin{tabular}{@{}ccccc@{}}
\toprule
Simulation & Number of tokens & First epoch best & Unique valid state ($u$) & ${\bar{u} \pm \sigma}$ \\ 
\midrule
1          & 4               & 236             & 67                      & \multirow{3}{*}{43$\pm$17} \\
2          & 4               & 631             & 36                      &                          \\
3          & 4               & 2920            & 27                      &                          \\ 
\midrule
4          & 5               & 304             & 18                      & \multirow{3}{*}{103$\pm$59} \\
5          & 5               & 401             & 145                     &                          \\
6          & 5               & 1332            & 145                     &                          \\ 
\midrule
7          & 6               & 5055            & 533                     & \multirow{3}{*}{445$\pm$74} \\
8          & 6               & 544             & 352                     &                          \\
9          & 6               & 1464            & 451                     &                          \\ 
\bottomrule
\end{tabular}
\end{table}

\begin{figure}[!htbp]
  \centering
  \includegraphics[width=0.7\linewidth]{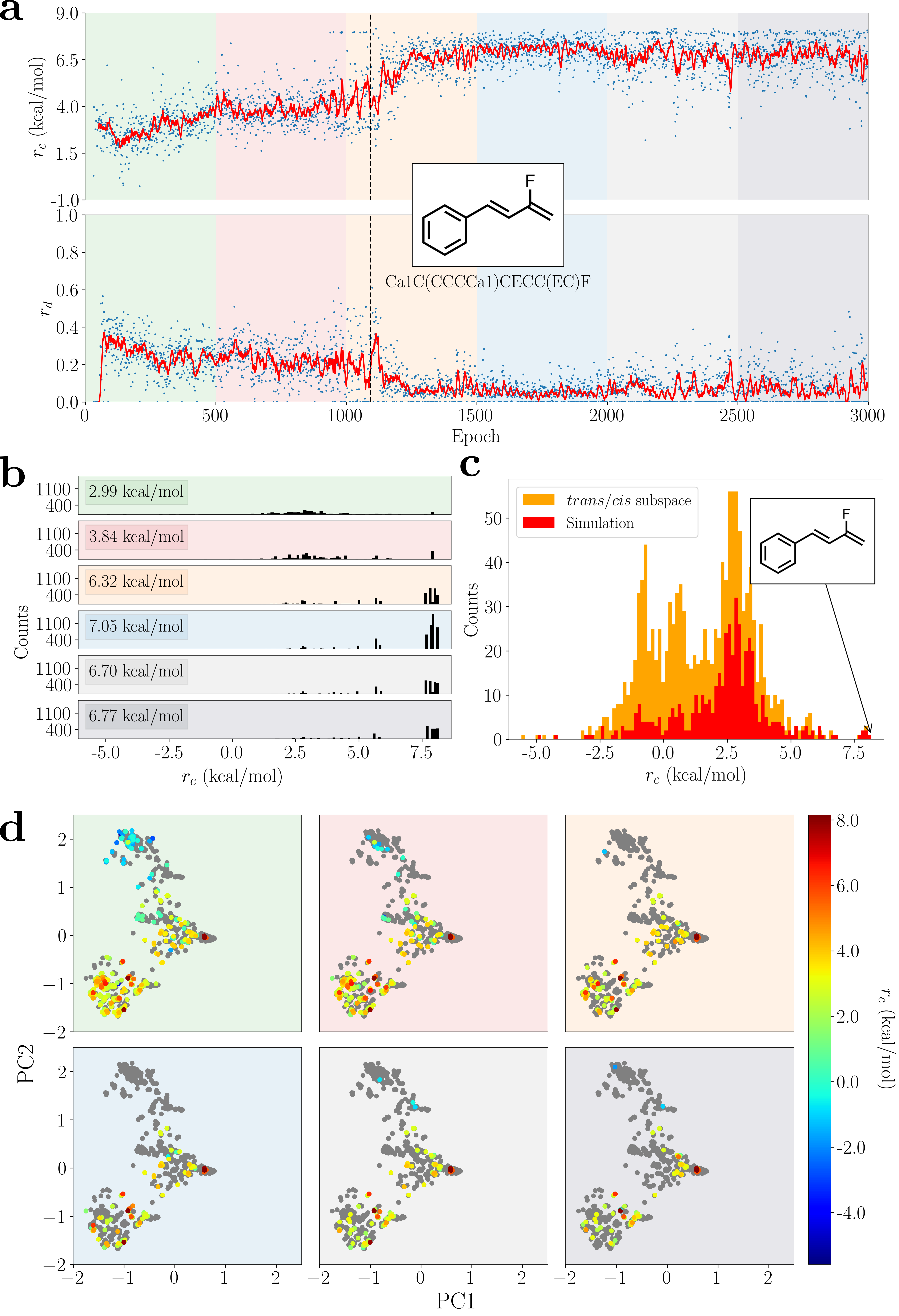}
  \caption{\textbf{Inverse design of \textit{trans}/\textit{cis} isomers with PROTEUS.} \textbf{a}, Time-evolution of the chemical reward during the PROTEUS simulation for the \textit{trans}/\textit{cis} isomers within the 6-token CS. Both the mean value of each epoch (blue scatter) and the running average (solid red line) are reported. The epoch corresponding to the first generation of the best solution (see the molecular formula in the inset), as ranked in the \textquoteleft \textit{trans}/\textit{cis} dataset', is marked with a dashed line. The 3,000 epochs reported are divided into 6 windows with different background colors. \textbf{b}, The counts of the generated molecules are plotted as a function of the chemical reward (i.e. the \textit{trans}/\textit{cis} energy gap), with mean values reported in the insets. \textbf{c}, The \textit{trans}/\textit{cis} energy gap distributions for unique isomers in the \textquoteleft \textit{trans}/\textit{cis} dataset' (in orange) and in the PROTEUS simulations (in red) are compared. \textbf{d}, For each simulation window, the generated molecules are displayed using the principal components defined for the full E/Z space (and reported in Fig. \ref{fig:panel1}c). The states generated in each window are labeled using the \textit{trans}/\textit{cis} energy gap values defined by the color bar, while molecules belonging to the \textit{trans}/\textit{cis} CS that are not explored are reported in grey.}
\label{fig:panel3}
\end{figure}

\FloatBarrier
\section{Supporting Information}\label{sec:secA2}

\subsection{SMILES and P-SMILES syntaxes} \label{sec:psmiles}
Chemoinformatics is a branch of computational chemistry and informatics that leverages informatic tools to process and work with chemical objects. One of the focuses of chemoinformatics is defining efficient ways to describe (and post-process) chemical objects using a computer. 
Among many, one of the most popular encodings is based on ASCII strings \cite{KRENN2022100588}. More precisely, molecules are defined as sequences of tokens that embed both chemical and structural features. 
Each token can be compared to a syllabus or a punctuation mark in natural languages since combinations of tokens return informative strings that convey the chemical properties of a system.

\subsubsection{From SMILES to P-SMILES. General properties}
SMILES is an extremely useful and complete syntax to describe chemical systems \cite{smiles1, smiles2, smiles3}. SMILES comprises letters, numbers, symbols, and their combinations. SMILES characters can be ideally classified as either \textit{content} or \textit{structural}. 

Content characters are the symbols of the corresponding element. However, since hydrogen atoms are implicitly considered, the valence of each non-hydrogen atom is always assumed as saturated by H atoms. For example, in the SMILES syntax \texttt{C} means methane, while \texttt{CC} means ethane. 
Content characters follow a double syntax and can be written in either capital or small letters. When an element is written in capital letters (e.g. \texttt{C}), the valence of its atom is saturated, i.e. belongs to an aliphatic system (unless followed by other specific characters), while when is written in small letters (e.g. \texttt{c}), its valence is unsaturated to an aromatic system. This turns into the equivalency between \texttt{c} and \texttt{C=} or \texttt{=C}, where \texttt{=} is a structural content for double bonds.

Structural characters can be classified based on the type of structural information encoded. They define, for example,  the order of a bond (i.e. \texttt{=} if double and \texttt{\#} if triple), the relative placement of moieties around a double bond (e.g. \texttt{X/C=C/Y} to define the E isomer of a \texttt{C=C} bond, and \texttt{X/C=C\char`\\ Y} to define the Z isomer of a \texttt{C=C} bond), branches, and cycles. Branches and cycles are embedded in pairs of brackets and numbers, respectively. For example, in SMILES syntax tertbutanol is encoded as \texttt{CC(C)(C)O}, while methy clcyclohexane is \texttt{CC1CCCCC1}. 

Aromatic systems in SMILES naturally feature equivalent definitions, e.g. benzene can be written as either \texttt{c1ccccc1} or \texttt{C1=CC=CC=C1}. In the context of RL generation where each action defines the addition of one (or multiple) tokens, in the absence of extensive pretraining, both chemically equivalent strings for benzene are very challenging for generative models since require a long sequence of juxtaposed tokens to be correct. Moreover, in SMILES, aromatic rings are \textit{special cases} in which the modification of just a single character (e.g. \texttt{=} in \texttt{C1=CC=CC=C1}) can break the aromaticity. This is not the same for the aliphatic rings that usually require more than one string modification to be converted into aromatic ones. 
This makes the generation of aromatic rings more unlikely than aliphatic ones, and more than how it is naturally supposed to be since an aromatic ring is a \textit{special} aliphatic ring. 

Similarly, the generation of an E isomer is more likely than for a Z one. In fact, the E isomer of a double bond can be written using three notations - i.e. \texttt{X=Y}, \texttt{/X=Y/}, and \texttt{\textbackslash X=Y\textbackslash} - while the Z isomer only two, i.e. \texttt{/X=Y\textbackslash} and \texttt{\textbackslash X=Y/}.
Notably, SMILES syntax defines \texttt{=} as a double bond with \textit{undefined stereochemistry}. However, in practice, when a SMILES string with a double bond with undefined stereochemistry is translated to Cartesian coordinates, it is considered an E isomer by default in several python libraries that implement SMILES \cite{obabel, pybel}. 

The inequalities in the syntactic complexity of aromatic/aliphatic rings and E/Z double bonds are a possible source of bias in RL-based generative models. Therefore, to leverage the probability of generating those features, we modified the SMILES syntax making it more PROTEUS-oriented (P-SMILES syntax). 
The idea behind P-SMILES is to reduce as much as possible the number of characters, i.e. the possible actions in the model, as well as the number of digits needed to define a syntactic meaning. 
As reported in Tab. \ref{table:psmiles}, we condensed the definition of E and Z isomers substituting the three-character notation of SMILES with a one-character one, i.e. introducing the \texttt{E} and \texttt{Z} tokens. 
Regarding the aromatic rings, we avoided the use of the small-letter notation for chemical elements and introduced a new class of cycles. The original SMILES notation relying on couples of numbers is joined by pairs of numbers marked with an \texttt{a}, e.g. \texttt{a1}, that define aromatic rings. The new token is specific for aromatic rings and leverages the syntactic complexity of the aliphatic analogs. In fact, if the \texttt{Ca1CCCECCa1} P-SMILES is generated, it is promptly converted to benzene, giving priority to the \texttt{a1} token rather than the \texttt{E} one.

Overall, PROTEUS reduces the number of symbols used in SMILES, while keeping its encoding capabilities. To summarize, the vocabularies of tokens to define carbon, nitrogen, oxygen, fluorine, single, double, and triple bonds, branches, and rings are:
\begin{enumerate}
    \item SMILES: $[$\texttt{C}, \texttt{N}, \texttt{O}, \texttt{c}, \texttt{n}, \texttt{o}, \texttt{F}, \texttt{/}, \texttt{\textbackslash}, \texttt{=}, \texttt{\#}, \texttt{1}, \texttt{(}, \texttt{)}$]$,
    \item P-SMILES: $[$\texttt{C}, \texttt{N}, \texttt{O}, \texttt{F}, \texttt{E}, \texttt{Z}, \texttt{\#}, \texttt{1}, \texttt{a1}, \texttt{(}, \texttt{)}$]$.
\end{enumerate}

\begin{table}[!htbp] 
\centering
\caption{\textbf{The P-SMILES syntax. }Representative differences between the SMILES and the P-SMILES syntaxes.}\label{table:psmiles}
\begin{tabular}{@{}ccc@{}}
\toprule
\textbf{Property} & \textbf{SMILES} & \textbf{P-SMILES} \\ [0.5ex]
\midrule
E conformation & \texttt{/X=Y/} or \texttt{\char`\\ X=Y\char`\\} or \texttt{X=Y} & \texttt{X\textbf{E}Y} \\
Z conformation & \texttt{/X=Y\char`\\} or \texttt{\char`\\ X=Y/} & \texttt{X\textbf{Z}Y} \\
Aromatic ring  & \texttt{C1=CC=CC=C1} or \texttt{c1ccccc1} & \texttt{C\textbf{a1}CCCCC\textbf{a1}} \\
\bottomrule
\end{tabular}
\end{table}

\FloatBarrier
\subsubsection{Performances of P-SMILES syntax}
To verify the performances of the P-SMILES syntax, we carried out a comparative study between P-SMILES and SMILES. For P-SMILES, we imposed that branching characters, i.e. $[$\texttt{1}, \texttt{a1}, and \texttt{( )}$]$, are always complete. This means that branching characters appear always in pairs. Such a constraint mimics the actual generation played by PROTEUS, which is designed to always choose a complete set of those tokens that are meaningful in juxtaposed pairs.

First, we compared the expressiveness of the SMILES and P-SMILES syntaxes. To do this, we benchmarked the ratios between E and Z double bonds and the ratio between aliphatic and aromatic rings in strings randomly sampled for each syntax. All the strings sampled are composed of a random number of tokens comprised between 8 and 20. All the values reported below are averaged over 20 independent batches, each composed of 25,000 samples. The vocabularies used for this test are:
\begin{enumerate}
    \item $[$ \texttt{\textbackslash}, \texttt{/}, \texttt{=}, \texttt{1}, \texttt{C}, \texttt{c} $]$ for SMILES,
    \item $[$ \texttt{E}, \texttt{Z}, \texttt{a1}, \texttt{1}, \texttt{C} $]$ for P-SMILES.
\end{enumerate}

\begin{figure}[!htbp]
\centering
    \includegraphics[width=.6\linewidth]{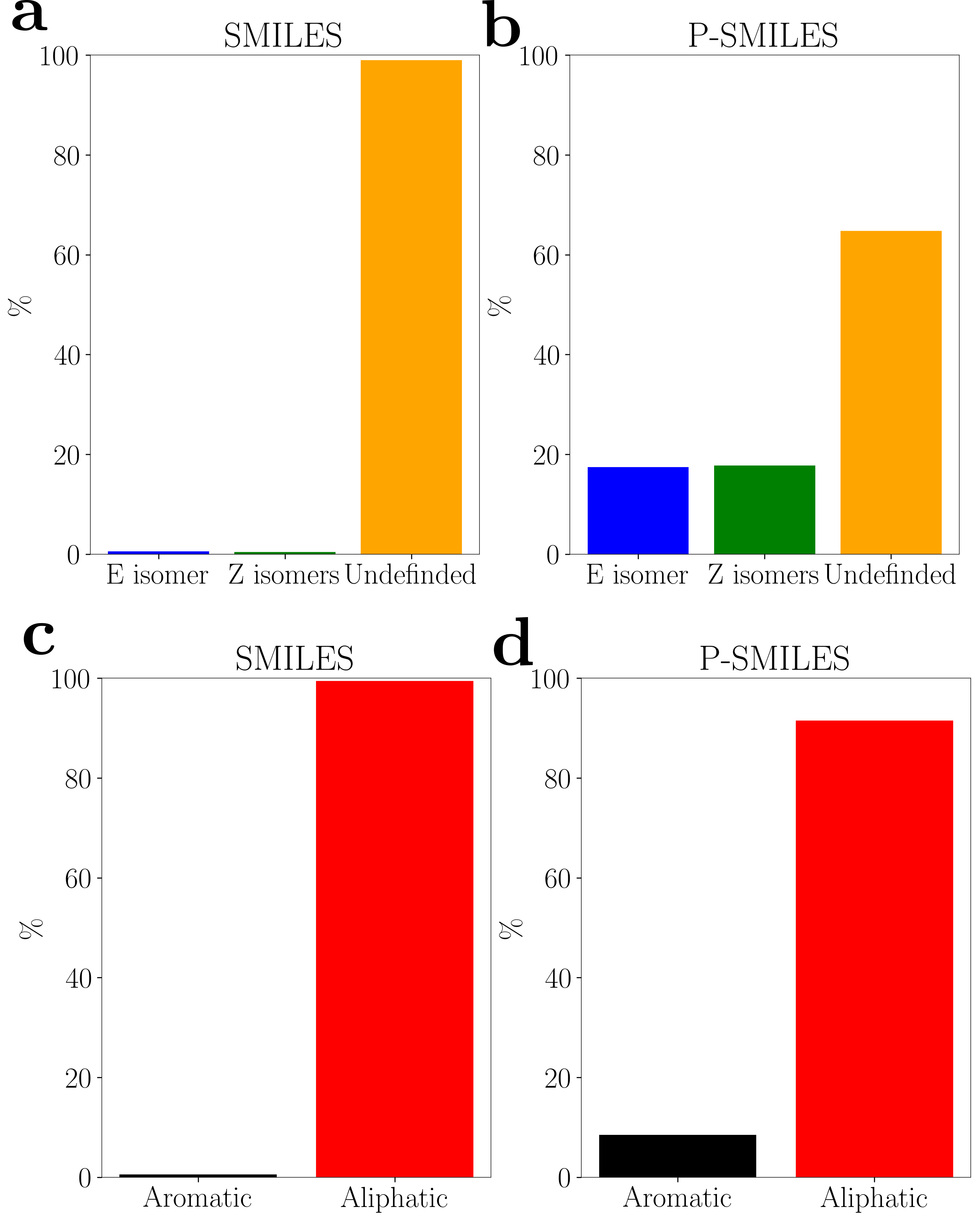}
\caption{\textbf{P-SMILES vs. SMILES.} Relative probabilities of building a double bond with E or Z configuration or an aliphatic or an aromatic ring using (\textbf{A} and \textbf{C}) SMILES or (\textbf{B} and \textbf{D}) P-SMILES syntax.}
\label{fig:aliaro}
\end{figure}

As shown in Fig. \ref{fig:aliaro}, the SMILES syntax favors double bonds with undefined stereochemistry. Double bonds with undefined stereochemistry are 99.0\% of the total, while well-defined E and Z double bonds are 5.7\% and 4.3\%, respectively. This means that stereochemically labeled double bonds are evenly generated, with a preference for E isomers.
When the P-SMILES syntax is turned on, the amount of double bonds with defined stereochemistry increases up to 17.4\% for the E isomers and to 17.8\% for the Z ones, while the number of double bonds without any specified stereochemistry is 64.8\% (Fig. \ref{fig:aliaro}A and \ref{fig:aliaro}B). The fact that there is a significant number of undefined double bonds is meaningful since all double bonds that are not isomeric (e.g. ketones) cannot have any stereochemical label. However, the fact that this number decreases suggests that within the set of double bonds with undefined stereochemistry in SMILES, several are E isomers.

Aromatic rings are rarer than aliphatic ones by definition since a ring is aromatic if it follows the H\"uckel's rule. However, generating an aromatic ring using SMILES is rarer than it should be. As reported in Fig. \ref{fig:aliaro}C, within the SMILES syntax the probability that in a random string, a ring is labeled as aromatic is close to zero (0.54\%). Instead, using the P-SMILES syntax, the probability of generating an aromatic ring rises to 8.7\% (Fig. \ref{fig:aliaro}D).

These outcomes point out the fundamental issue of the bias within the SMILES syntax and, thus, its limitations in RL simulations. Since the isomeric configuration and aromaticity labeling are biased by the use of complicated (and often unequal) combinations of tokens, the exploration of some regions of the space of solution could be limited, even in the limit of using a highly representative pretrained language model. 
The P-SMILES syntax we developed is more suitable for RL applications since it ideally levels out syntax complexities while limiting the maximum number of tokens needed to define structural isomers and aromatic rings up to two.

\FloatBarrier
\subsection{Implementation details}  

\subsubsection{System architecture}
All the models share an initial transformer encoder: given an input sequence, the encoder embeds the sequence of tokens in continuous vectors, adds a positional encoding to characterize the relative position of different tokens, and processes them through a multi-head self-attention layer \cite{vaswani2017}. The encoder outputs a continuous vector for each token, that is the input information for subsequent the agent's models. 

The master, $\pi_{M}\!\left(s_t\right)$, receives as input a state $s_t$ and decides among three actions: 
\begin{enumerate}
    \item \texttt{add single-character token}
    \item \texttt{add double-character token}
    \item \texttt{return $s_t$}.
\end{enumerate}
If the first action is selected, $s_t$ is passed to the single-character position predictor, $\pi_{P^S}\!\left(s_t\right)$, which returns a vector of probabilities related to all the possible positions. In the returned position index $i$ of $s_t$, a \texttt{[MASK]} token is added. Next, the modified $s_t$ is passed to the single-character generator, $\pi_{G^S}\!\left(s_t\right)$, which returns a vector of probabilities related to all the possible single-character tokens. Finally, a token $v$ is sampled and used to replace \texttt{[MASK]} in $s_t$ and get $s_{t+1}$. 
If the \texttt{add double-character} action is selected, $s_t$ is passed to the double-character position predictor, $\pi_{P^D}\!\left(s_t\right)$, which returns two vectors of probabilities, one for each position to be chosen. Two indexes $i_0$ and $i_1$ are sampled so that $i_0 :< i_1$. After inserting two \texttt{[MASK]} tokens to $s_t$, each at indexes $i_0$ and $i_1$, the double-character generator, $\pi_{G^D}\!\left(s_t\right)$, decides which pair of characters will be inserted by returning a vector of probabilities related to all the possible pairs of characters. The sampled pair $v_0$, $v_1$ is then used to replace \texttt{[MASK]} tokens and get $s_{t+1}$. 
Finally, if the \texttt{return $s_t$} action is selected, the generation is considered as concluded. 
The complete pseudocode for the generative loop is provided in Algorithms \ref{alg:gen}. 

\begin{figure}[!htbp] 
    \centering
    \includegraphics[width=\textwidth]{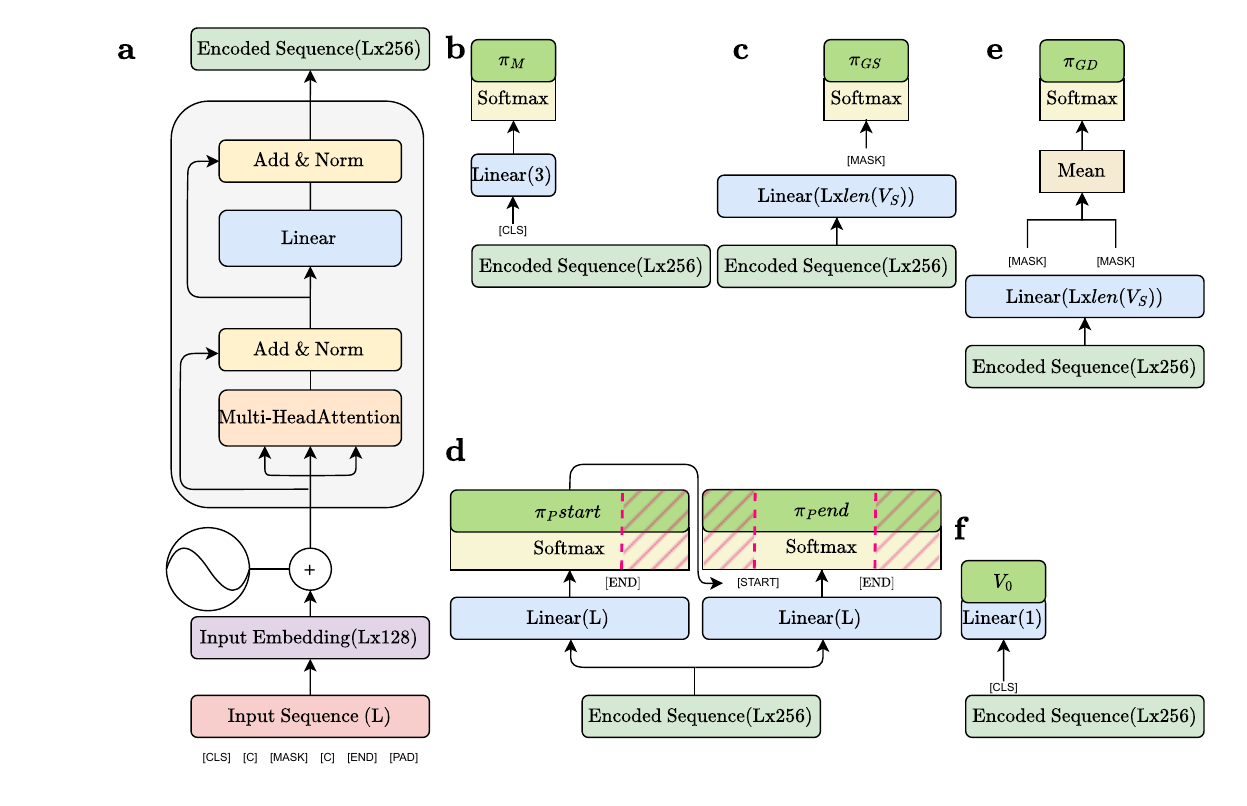}
    \caption{\textbf{PROTEUS architecture.} The overall architecture of our system. \textbf{a}, The encoder. \textbf{b}, The master, $M$. \textbf{c}, The single-character generator, $G^S$. \textbf{d}, The two position predictors, $P^D$ and $P^S$. \textbf{e}, The double-character generator, $G^D$. \textbf{f} , The critic. The input sequence at the token level is marked in red; the encoded sequence as a sequence of continuous vectors is displayed in green; the outputs of the models as probability distributions are reported in lime; the linear layers are highlighted in blue; non-trainable mathematical functions are in yellow; embedding layers are reported in purple, and multi-head attention in orange. The sizes of inputs and NN layers are reported in brackets, with L the length of the generated P-SMILES, i.e. $s_t$.}
    \label{fig:fullmodel}
\end{figure}

As pictorially depicted in Fig. \ref{fig:fullmodel}, the master uses a single fully connected layer with softmax activation to return the probability of its three possible actions. The two position predictors share a fully connected layer with softmax activation to return the probabilities of the first position, while only the double-character position predictor uses also a second layer to predict the softmax probabilities of the second position. The single-character generator starts from the encoding of the target token and applies a fully connected layer with softmax activation to get the probabilities of the single characters. Similarly, the double-character generator starts from the encoding of the two target tokens, applies a fully connected layer, and then averages their outputs before applying the softmax activation to return the probabilities for the double characters. Finally, the critic takes the initial encoded sequence and returns the estimate of the value of the current state.

\FloatBarrier
\subsubsection{Pseudocode of PROTEUS}
The pseudocode for generating P-SMILES with our method is reported in Algorithm \ref{alg:gen}. Instead, the pseudocode for the complete training of PROTEUS models is available in Algorithm \ref{alg:training}.

\begin{algorithm}[!htbp] 
\caption{P-SMILES generation with PROTEUS}\label{alg:gen}
\begin{algorithmic}[1]
\Require $\pi_{M}$, $\pi_{P^D}$, $\pi_{G^D}$, $\pi_{P^S}$, $\pi_{G^S}$, max length $L$
\State \textbf{initialize} $s = \text{empty\_vector}$
\Repeat 
    \State $a \sim \pi_{M}(s)$
    \If{$a$ is \texttt{add double-character token}}
        \State $i_0, i_1 \sim \pi_{P^D}(s)$
        \State insert \texttt{[MASK]} tokens to $s_{i_0}$ and $s_{i_1}$
        \State $v_0, v_1 \sim \pi_{G^D}(s)$
        \State $s_{i_0}, s_{i_1} = v_0, v_1$
    \ElsIf{$a$ is \texttt{add single-character token}}
        \State $i \sim \pi_{P^S}(s)$
        \State insert \texttt{[MASK]} token to $s_i$
        \State $v \sim \pi_{G^S}(s)$
        \State $s_i = v$
    \EndIf
\Until{$a$ is of termination \textbf{or} $|s| \geq L$}
\State \textbf{return} $s$
\end{algorithmic}
\end{algorithm}

\begin{algorithm}[!htbp] 
\caption{The training procedure of PROTEUS}\label{alg:training}
\begin{algorithmic}[1]
\Require Backbone $b$, batch size $B$, memory size $n$, max sequence length $L$, scaling factors $\alpha$ and $\beta$
\State \textbf{initialize} $\pi_{M}$, $\pi_{P^D}$, $\pi_{G^D}$, $\pi_{P^S}$, $\pi_{G^S}$, $D$, top-K
\While{stop condition not satisfied}
    \State Generate $B$ P-SMILES ($S$) via Algorithm \ref{alg:gen}
    \State Append $b$ to $s$ $\forall s \in S$
    \State Check validity of $s$ $\forall s \in S$
    \ForAll{$s$}
        \If{$s$ is valid}
            \State Compute fitness $r_c$ of $s$ (see Sec. \ref{sec:fitness})
            \State append $(s, r_c)$ to top-K
            \State sort top-K based on $f$
            \State pop last element from top-K
            \State Compute diversity $r_d$ of $s$ (Equation \ref{eq:diversity})
            \If{$s \in D$}
                \State pop $s$ from $D$
                \State append $s$ to $D$
            \ElsIf{$s \notin D$}
                \State pop last from $D$ if $|D| = n$
                \State append $s$ to $D$
            \EndIf
            \State $r_t = \alpha r_c(s_t) + \beta r_d(s_t)$
        \ElsIf{$|s| = 0$ or $|s| > L$}
            \State $r_t = -1$
        \Else
            \State $r_t = 0$
        \EndIf
    \EndFor
    \State Backpropagate to models prioritizing top-K molecules using PPO (Equation \ref{eq:ppo})
\EndWhile
\State \textbf{return} top-K
\end{algorithmic}
\end{algorithm}

\FloatBarrier
\subsubsection{Computational resources and parallelization implementation} \label{sec:comp_resources}

The PROTEUS simulations follow a complex routine with two main computational bottlenecks: 
\textit{i}) the training of the NNs, which requires high-performing GPU hardware, and \textit{ii}) calculations to compute $r_c$ (see Sec. \ref{sec:methods} in the main text for details), which require performing CPU hardware. Furthermore, QM software like xTB \cite{xtb}, CREST \cite{crest} and Gaussian16 \cite{g16} are often run in multi-thread systems. All the calculations were done using a hardware setup with 96 CPUs (1 thread each) Intel(R) Xeon(R) Gold 6252N CPU @2.30GHz and 2 GPUs NVIDIA Tesla T4 (CUDA Version: 12.1, Driver Version: 530.30.02).

To lower the impact of the cost of QM calculations, the routine to compute the $r_c$ is parallelized. Since PROTEUS generates a variable number of valid P-SMILES strings each epoch,  firstly steps from 1 to 3 in Sec. \ref{sec:fitness} are executed for all the samples in a generated batch. In this way, syntactically-valid states are filtered at a low computational cost. Next, according to the number of threads allocated for the simulation, PROTEUS automatically assigns the maximum number of threads to each isomer and executes steps from 4 to 10 in Sec. \ref{sec:fitness} in parallel for each candidate molecule. Once all the routines running in parallel are completed, PROTEUS executes step 11 (Sec. \ref{sec:fitness} in the main text) to compute an array containing the $r_c$ value of each state within the given epoch. 
In this way, the maximum computational parallelization to compute $r_c$ is always ensured.

\FloatBarrier
\subsection{Insights into the \textquoteleft E/Z dataset'} \label{sec:insEZdb} 

The number of possible P-SMILES string combinations for the [\texttt{E}, \texttt{Z}, \texttt{a1}, \texttt{1}, \texttt{\#}, \texttt{(}, \texttt{)}, \texttt{C}, \texttt{N}, \texttt{O}, \texttt{F}] vocabulary comprising up to 6 tokens is 1,948,716, which means 1,948,716 pairs of isomers. 273,718 are syntactically complete and valid pairs. Within this subset, 2,156 pairs of isomers are neutral and closed-shell molecules. Therefore, the number of $r_c$ computations is 4,312. In the most-accurate energy evaluation routine, i.e. when the most stable conformer sampled by CREST is optimized at DFT level, 1,056 did not pass all the validity tests: 
\begin{enumerate}
    \item 4 strings did not be converted to the corresponding Cartesian coordinates, due to OpenBabel inconsistency, i.e. \texttt{Ca1C(CCCCa1)/C=C/c1\#cn1}, \texttt{and Ca1C(CCCCa1)/C=C\textbackslash c1\#cn1}, \texttt{Ca1C(CCCCa1)/C=C/c1\#nc1}, and \texttt{Ca1C(CCCCa1)/C=C\textbackslash c1\#nc1}.
    \item 119 molecules experienced errors or changes in the number of bonds during the first xTB optimization.
    \item 21 molecules were not fully optimized by Gaussian16, mainly due to convergence issues. 
    \item 912 molecules experienced other kinds of changes in the connectivity at the end of the calculations.
\end{enumerate}
Therefore, the final dataset contains a total of 1,628 pairs of isomers of the styrene backbone for which the E/Z energy gap was computed correctly.
1,246 out of 1,628 pairs of molecules feature the R$^2$ substituent which is a hydrogen atom.

When the maximum length of the generated P-SMILES is increased by just one token raises the number of syntactically valid closed-shell pairs of \textit{trans}/\textit{cis} isomers to be tested with QM methods from 1,644 to 6,005. Considering the estimated cost of the QM-based validity test, comprising a rigorous conformational sampling (Fig. \ref{fig:panel5}), for the 4-, 5-, and 6-token CSs and the additional cost introduced by including an extra token, i.e. involving larger and more flexible molecules, the computational effort to fully characterize the 6,005 pairs of 7-token CS becomes extremely demanding. 

\FloatBarrier
\subsubsection{Benchmark of the reward function}
\begin{figure}[!htbp] 
  \centering
  \includegraphics[width=0.8\linewidth]{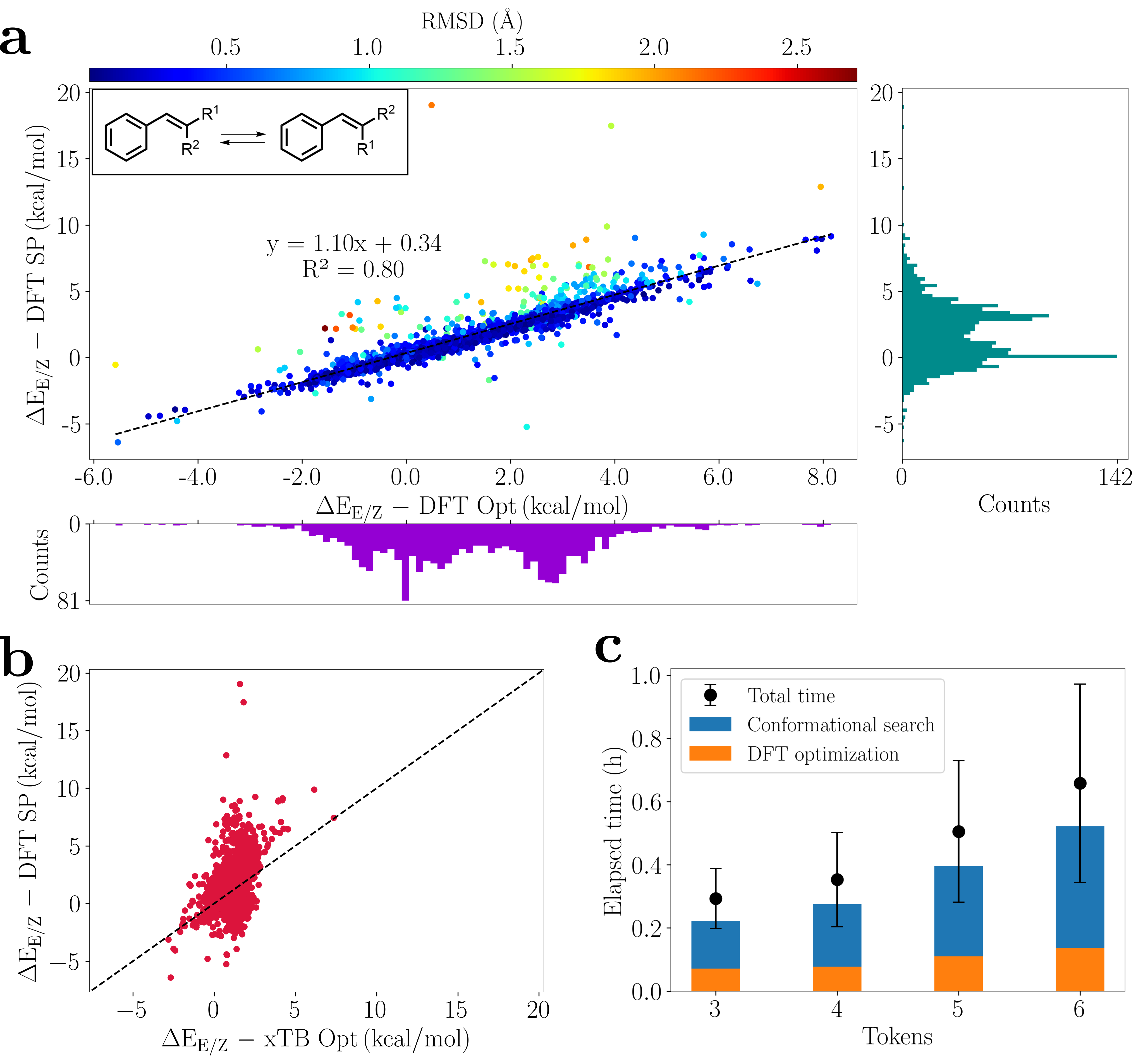}
  \caption{\textbf{Energy correlation and computational cost.} \textbf{a}, Linear regression analysis of the E/Z energy gaps in the reference \textquoteleft E/Z dataset' computed by single-point DFT calculations of DFT-TB optimized geometries and DFT geometry optimizations. Color coding is based on the maximum root mean square deviation (RMSD) between DFT-optimized and non-optimized geometries. The distributions of E/Z energy gaps are reported for both single-point calculations (light cyan histogram, right panel) and DFT-optimized geometries (dark violet histogram, bottom panel). \textbf{b}, Correlation between energy gap computed with GFN-xTB and B3LYP/6-31G(d,p) single-point on top of the conformational search for molecules in the reference \textquoteleft E/Z dataset'. \textbf{c}, Average time for the conformational search calculations with the CREST software (blue) and the DFT optimization with the Gaussian16 software (orange) for different numbers of tokens using 6 CPUs. Results for only valid molecules in the reference \textquoteleft E/Z dataset' are reported.}
  \label{fig:panel5}
\end{figure}

The final electronic energy of the most stable conformer according to CREST of the molecules generated by PROTEUS was computed at three different levels: \textit{i}) DFT-TB optimized geometries, \textit{ii}) DFT single-point calculations on top of \textit{i}), or \textit{iii}) DFT optimized geometries. The isomeric energies with each method were compared (Fig. \ref{fig:panel5}). Since DFT is a higher level of QM theory, energies of DFT-optimized geometries were used as reference.

First, we compared the DFT-TB and single-point DFT energies. As shown in Fig. \ref{fig:panel5}b, the correlation is poor. 
Moreover, the energies computed using DFT-TB range between -1 and 3 kcal/mol ca., while the DFT energies are distributed in a much larger range, i.e. between -5 and 8 kcal/mol.
Further analyses gave insights into the difference between DFT energies computed for geometries optimized with DFT-TB and DFT. Namely, we investigated both the quality of DFT-TB geometries and, thus, their relative single-point DFT energies. As reported in Fig. \ref{fig:panel5}a, energies are almost linearly correlated (\textit{R}$^2$=0.80). 
Calculations also showed that the correlation is better when the maximum root mean square deviation (RMSD) value between the geometry of DFT- and DFT-TB-optimized geometries is little, i.e. ca. <0.50 \AA. On the contrary, molecules that feature large energy differences before and after the DFT optimization generally feature large RMSD values (> 1.25 \AA). For a few geometries, the single-point calculation overestimates the energy gap by more than 15 kcal/mol (Fig. \ref{fig:panel5}a). Even if rare, such a behavior is potentially very dangerous for the learning process, since it would lead to over-rewarding some molecules and introduce bias. 

To assess the overall computational cost of QM calculations, which are the actual computational bottleneck in PROTEUS, the average time required for CREST simulations and DFT optimization was analyzed. Fig. \ref{fig:panel5}c shows that the biggest effort is required for the indispensable conformational analysis. CREST calculations require more than double the time of DFT optimizations. In the case of the 6-token subset of molecules, the computational cost to sample the conformers is ca. 5 times higher than the DFT optimization (Fig. \ref{fig:panel5}c). Not surprisingly, the total computational cost increases in function of the length of P-SMILES strings, since the number of the molecular degrees of freedom increases as well, making the conformational space to explore wider.

\FloatBarrier
\subsubsection{About the impact of the QM method on the E/Z energy gap } \label{sec:impact_solvation}

It is crucial to note that targeting high accuracy in the computation of the isomerization energies is not really relevant for developing an appropriate architecture of PROTEUS. In this work, in fact, we aim at an ML tool able to explore broadly large CSs and exploit properly the chemical property associated with the involved molecular states, independently from the specific values of the chemical rewards and, thus, from the level of QM method used. 
In fact, the generation capability of PROTEUS is conditioned by its ML architecture, while the quality of the generated molecules depends on the level of theory chosen for computing $r_c$. This choice is a key step in any computational chemistry study and is strictly related to the final target of the work. 
For example, the absolute and relative energy values are changed whenever dispersion, entropic, and solvation effects are included or the basis set is varied. In the present section, the effect of these corrections is described for the best and worst 5 molecules as ranked in the reference \textquoteleft EZ dataset'. 

On top of the geometry optimized with DFT, the final energy was refined with the following protocols: 
\textit{i}) the thermal correction was computed at room temperature with B3LYP/6-31G(d,p), 
\textit{ii}) the basis-set correction was included by computing the single-point energy at the B3LYP/6-311++G(2d,2p) \cite{bb1,bb2,bb3,bb4} level of theory, 
\textit{iii}) the effect of the dispersion was calculated by computing the single-point energy at the B3LYP-D3(BJ)/6-311++G(2d,2p) \cite{d3,BJ} level of theory, 
iv) the effect of the solvation correction was estimated by computing the single-point energy at the B3LYP-D3(BJ)/6-311++G(2d,2p)/SDM(water) \cite{SDM} level of theory.

As reported in Tab. \ref{tab:benchmark-study}, the thermodynamics of the best and worst 5 molecules is consistent for all the methods, i.e. no changes in the sign of the energy gaps occur. However, the relative ranking is highly dependent on the computational protocol. This outcome is not surprising, since any change, e.g. the presence of an implicit solvent, may lead to non-negligible modifications in the local electron density resulting in an extra-stabilization of some isomers. 
For example, when dispersion and solvent corrections are included, the energy gap of \texttt{Ca1C(CCCCa1)CECC(EO)C} varies significantly (from 7.97 to 5.60 kcal/mol) likely due to the presence of the electron-withdrawing group C=O, which is sensitive to variations in the environment. Similarly, dispersion and solvation corrections reduce the energy gap.

Even if different computational protocols turn into different rankings of the same set of molecules, it is important to mark again that it does not affect the machinery of PROTEUS nor its capability to generate candidates that successfully maximize a property. 

\begin{table}[!htbp] 
    \centering
    \resizebox{\textwidth}{!}{
    \begin{tabular}{@{}ccccccccc@{}}
        \hline
        \textbf{\makecell[c]{P-SMILES\\(R$^1$/R$^2$)}} & $\mathbf{\Delta E_{\textbf{E/Z}}^{\textbf{\shortstack{B3LYP/\\SBS}}}}$ & $\mathbf{\Delta G_{\textbf{E/Z}}^{\textbf{\shortstack{B3LYP/\\SBS}}}}$ & $\mathbf{\Delta E_{\textbf{E/Z}}^{\textbf{\shortstack{B3LYP/\\BBS//\\B3LYP/\\SBS}}}}$ & $\mathbf{\Delta G_{\textbf{E/Z}}^{\textbf{\shortstack{B3LYP/\\BBS//\\B3LYP/\\SBS}}}}$ & $\mathbf{\Delta E_{\textbf{E/Z}}^{\textbf{\shortstack{B3LYP-D3(BJ)/\\BBS//\\B3LYP/\\SBS}}}}$ & $\mathbf{\Delta G_{\textbf{E/Z}}^{\textbf{\shortstack{B3LYP-D3(BJ)/\\BBS//\\B3LYP/\\SBS}}}}$ & $\mathbf{\Delta E_{\textbf{E/Z}}^{\textbf{\shortstack{B3LYP-D3(BJ)\\/BBS/SDM(water)//\\B3LYP/SBS}}}}$ & $\mathbf{\Delta G_{\textbf{E/Z}}^{\textbf{\shortstack{B3LYP-D3(BJ)/\\BBS/SDM(water)//\\B3LYP/SBS}}}}$ \\
        \toprule
        \multicolumn{9}{l}{\textbf{Top 5}}\\
        \texttt{C(EC)F} & 8.15 & 8.25 & 7.82 & 7.92 & 6.56 & 6.66 & 6.97 & 7.07 \\
        \texttt{C(EO)C} & 7.97 & 7.94 & 7.03 & 6.99 & 5.37 & 5.33 & 5.64 & 5.60 \\
        \texttt{CCENOC} & 7.96 & 10.09 & 8.72 & 10.85 & 6.51 & 8.64 & 4.79 & 6.92 \\
        \texttt{CCCENO} & 7.88 & 8.85 & 8.71 & 9.68 & 5.43 & 6.40 & 3.88 & 4.85 \\
        \texttt{C(EC)O} & 7.87 & 8.06 & 8.03 & 8.21 & 6.80 & 6.99 & 7.12 & 7.30 \\
        \midrule
        \multicolumn{9}{l}{\textbf{Worse 5}}\\
        \texttt{(NEO)O} & -4.44 & -4.10 & -4.04 & -3.71 & -4.12 & -3.78 & -3.48 & -3.15 \\
        \texttt{OCNOC} & -4.73 & -5.23 & -5.04 & -5.54 & -3.83 & -4.33 & -1.71 & -2.21 \\
        \texttt{OCNOCN} & -4.95 & -5.58 & -5.63 & -6.26 & -3.91 & -4.54 & -1.80 & -2.43 \\
        \texttt{(F)NOF} & -5.54 & -4.62 & -6.16 & -5.24 & -5.91 & -4.99 & -6.97 & -6.06 \\
        \texttt{ECCONZF} & -5.58 & -4.80 & -6.73 & -5.95 & -6.31 & -5.53 & -4.54 & -3.76 \\
        \bottomrule
    \end{tabular}}
    \caption{\textbf{Comparison of computational protocols.} Comparison of the effects of different quantum chemistry protocols in estimating the E/Z energy gap. SBS = 6-31G(d,p), BBS = 6-311++G(2d,2p).}
    \label{tab:benchmark-study}
\end{table}

\FloatBarrier
\subsubsection{Top and worst molecules in the \textquoteleft E/Z dataset'}
\begin{figure}[!htbp] 
    \includegraphics[width=1.\linewidth]{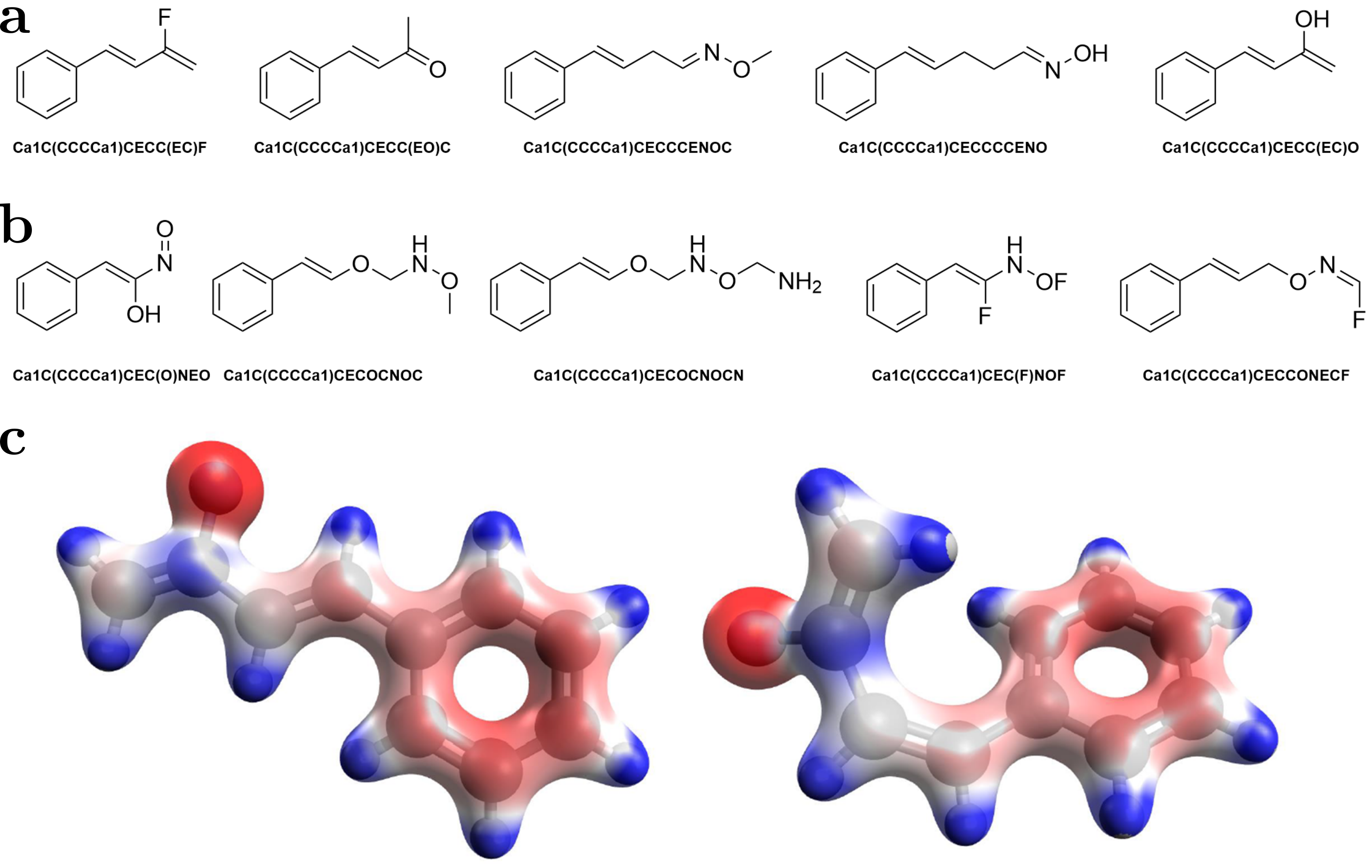}
\caption{\textbf{Top and worse molecules in the \textquoteleft E/Z dataset'.} Top-5 (\textbf{a}) and worse-5 (\textbf{b}) molecules in the \textquoteleft E/Z dataset', according to E/Z energy gap computed at B3LYP/6-31G(d,p) level of theory on geometries optimized with DFT. \textbf{c}, Electron densities colored by electrostatic potentials calculated at B3LYP/6-31G(d,p) level of theory for \texttt{Ca1C(CCCCa1)CECC(EC)F} (left side) and \texttt{Ca1C(CCCCa1)CZCC(EC)F} (right side). The electrostatic potential ranges from negative (red regions) to positive (blue regions), and is plotted using the Avogadro software \cite{avogadro}.}
\label{fig:electron_density}
\end{figure}

The top-5 and worse-5 in the \textquoteleft E/Z dataset' are reported in Figg. \ref{fig:electron_density}a and \ref{fig:electron_density}b. Interestingly, the first 5 molecules in the reference dataset feature the same R$^2$=H ligand, whereas differ for the R$^1$ one. It is interesting to notice that the first two molecules feature an extended $\pi$-conjugated system where the styrene moiety is connected to an aliphatic $\pi$-system. Namely, the presence of several resonance structures leads to an extra-stabilized E isomer and, thus, a large $\Delta_{E/Z}$ gap. Moreover, the top-ranked molecules feature electron-withdrawing groups, e.g. fluorine atoms and ketones, that further stabilize the E isomer.
For example, \texttt{Ca1C(CCCCa1)CECC(EC)F} shows a $\pi$-conjugated system over the whole molecule and a fluorine moiety placed close to the double bond of the styrene backbone. The fluorine enhances the conjugation of electrons, by behaving as an electron-withdrawing group (Fig. \ref{fig:electron_density}c) and stabilizing the E conformer, whereas in the Z isomer, it forces to place the terminal =CH$_2$ close to one C\textsubscript{aromatic}-H group, behaving as a steric hindering group. This reduces the conjugation of the system since the \texttt{C(EC)F} moiety is no longer laying on the phenyl plane, destabilizing this isomer (Fig. \ref{fig:electron_density}b). 
On the contrary, worse-ranked molecules show electron-donating groups, e.g. N- and O-based moieties (Fig. \ref{fig:electron_density}b), which increase the electronic density towards the styrene backbone and destabilize the E isomer.

\FloatBarrier
\subsubsection{Clustering methods} \label{sec:clustering}
The visualization of the chemical space is a key point for the interpretation of both data and the learning process. A well-established procedure in the field requires \textit{i}) encoding molecules (i.e. SMILES or P-SMILES strings) as bit vectors according to the Morgan fingerprint scheme \cite{rdkit}, and then \textit{ii}) reducing the dimension to improve their visualization and interpretation, since plotting \textit{n}-dimensional (with \textit{n}>3) data is challenging.

In the present work, three different methods are compared: \textit{i}) the principal component analysis (PCA) \cite{pca}, \textit{ii}) the T-distributed Stochastic Neighbor Embedding (t-SNE) \cite{tSNE, scikit-learn}, and \textit{iii}) the Uniform Manifold Approximation and Projection (UMAP) \cite{umap, umap-software}. Each tentative of clustering the \textquoteleft E/Z dataset' reported here was carried out on Z isomers (Morgan fingerprints with radius = 5 and 4,096-bit).

\begin{figure}[!htbp] 
    \centering
    \includegraphics[width=.55\linewidth]{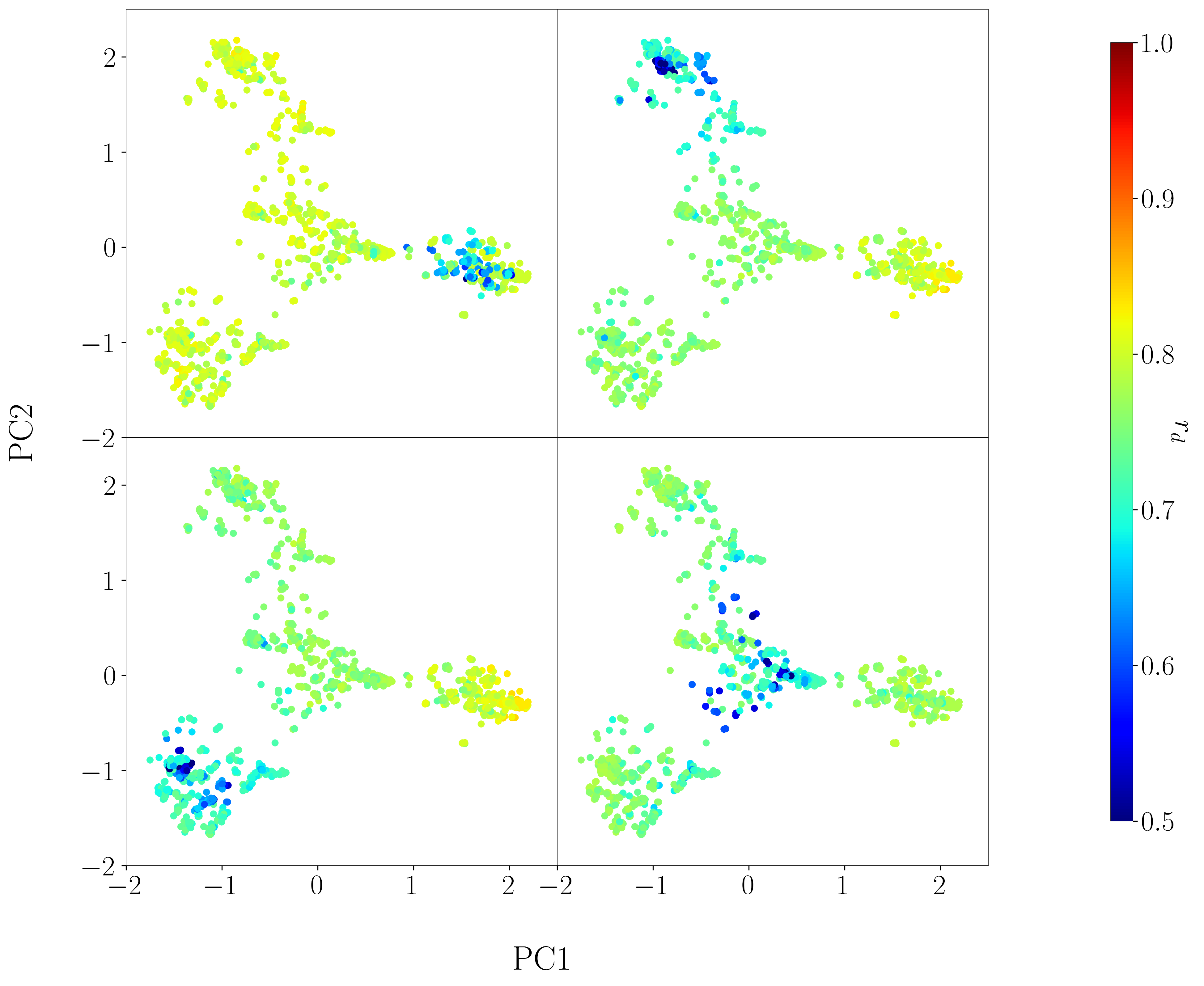}
    \caption{\textbf{PCA clusters and diversity value correlation.}  PCA clustering of the reference \textquoteleft E/Z dataset' based on Morgan fingerprint of Z molecules with color coding based on the diversity value computed for each molecule using a random molecule as reference. Molecules with a diversity value below 0.5 are labeled with a diversity value equal to 0.5.}
    \label{fig:PCA_clustering}
\end{figure}

PCA is a widely used methodology in several fields to reduce the dimension of data. The performance of PCA in reducing the number of dimensions is evaluated by computing the fraction of the overall variance recovered by the \textit{m} principal components chosen for the reduction. Ideally, this fraction should be close to 1, since it estimates how much of the original information is preserved after the transformation.
However, in the case of data with high dimensionality, like in the case of 4,096-bit Morgan fingerprints, the fraction of the overall variance recovered by the first two components is low, i.e. ca. 10\% in the present case. Therefore, PCA could appear as not the best tool to employ. On the contrary, PCA is an excellent visualization method for the \textquoteleft E/Z dataset' with better results than t-SNE and UMAP.
As already discussed in Sec. \ref{sec:problem} in the main text, the clustering of the dataset with PCA leads to four main clusters, which become three when the dataset is filtered by excluding those molecules with R$^2$=H  (Figg. \ref{fig:panel1}a and \ref{fig:panel1}b in the main text). This outcome clearly shows that a 2-component PCA already distinguishes a key structural characteristic of the molecules in the dataset, as the chemical species of R$^2$. Similarly, by plotting the diversity value between a random reference molecule and the remaining ones, it is clear that chemically similar molecules belong to the same cluster, while molecules that are chemically different usually belong to different clusters (Fig. \ref{fig:PCA_clustering}). As discussed in the main text, given the complex structure-property relationship investigated, the correlation between the clusters and the energy gap, in our specific case, is already gratifying. In fact, the quality of the relationship between the clusters and the property does not depend only on the performance of the clustering method, but mainly on the capability of the chosen encoding, i.e. the Morgan fingerprint, to correlate to the property itself. In plain words, the correlation is lacking if the encoding method does not contain any relevant information to describe the property. 

\begin{figure}[!htbp] 
  \centering
  \includegraphics[width=.8\linewidth]{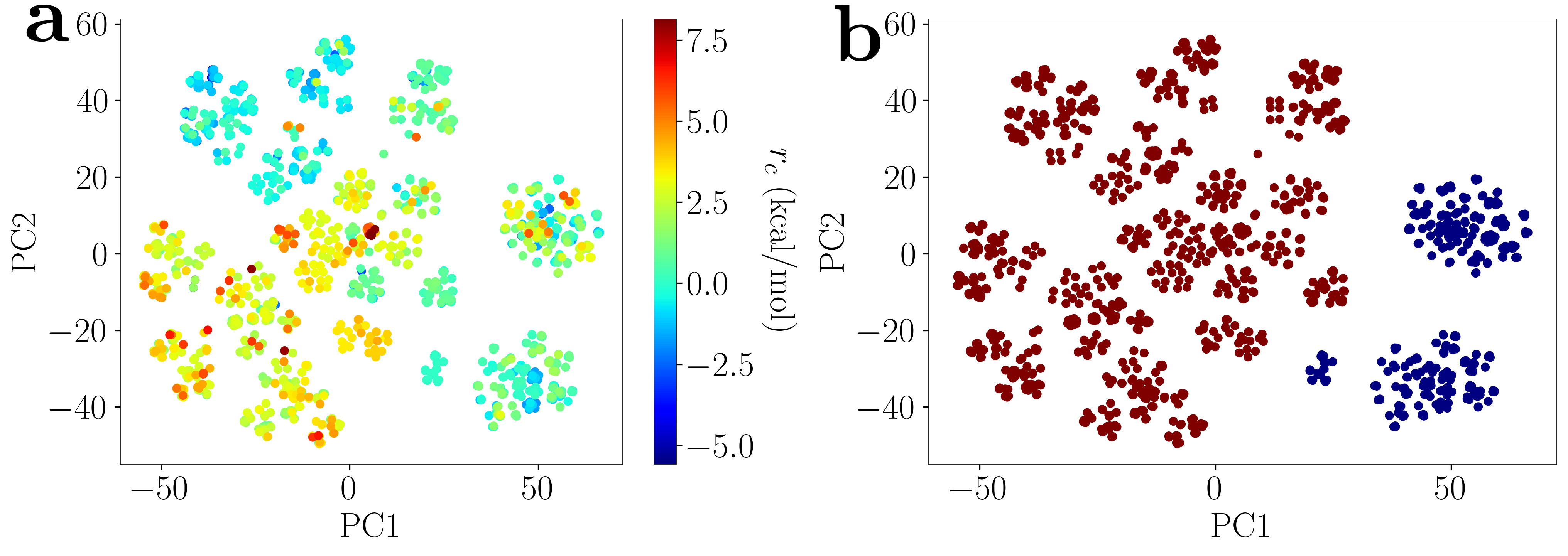}
  \caption{\textbf{t-SNE analysis.} Clustering of the reference \textquoteleft E/Z dataset' based on Morgan fingerprint of Z molecules and using the t-SNE algorithm. The clusters are colored according to (\textbf{a}) the energy gaps computed on DFT optimized geometries, and (\textbf{b}) the chemical structure of the R$^2$ moiety, i.e. dark red if R$_2 \ne $H and blue is R$^2$=H.}
  \label{fig:panel6}
\end{figure}

The performance of PCA was compared with the t-SNE method. Since t-SNE is a highly computationally demanding method, it is often used in conjunction with PCA. In fact, to save computational time, the number of dimensions of the original dataset is often reduced with an \textit{m}-components PCA choosing \textit{m} such that a significant part of the overall variance is recovered. Then, on top of PCA-transformed data, the t-SNE reduction is performed to improve the visualization of the \textit{l}-dimensions data, preserving the local structure of data.
Fig. \ref{fig:panel6}a shows the dimension reduction carried out with t-SNE on top of a PCA analysis done by choosing enough components to recover 95\% of the overall variance. The figure shows that t-SNE leads to more and better-shaped clusters than in PCA. However, similarly to the previous analysis the relationship between the clusters and the energy gap is not perfect and some clusters show both large and small energy gaps. 
Interestingly, even if 95\% of the overall variance is recovered, similar results are obtained as for the reference PCA. In fact, by coloring the t-SNE clusters in dark red or blue according to the conditions R$_2\neq $H or R$^2$=H, respectively, a clear separation between those two sets is highlighted as for PCA (Fig. \ref{fig:panel6}b). To summarize, it is noteworthy that the t-SNE reduction is less compact than PCA, without offering any further information.

\begin{figure}[!htbp] 
    \centering
    \includegraphics[width=1.\linewidth]{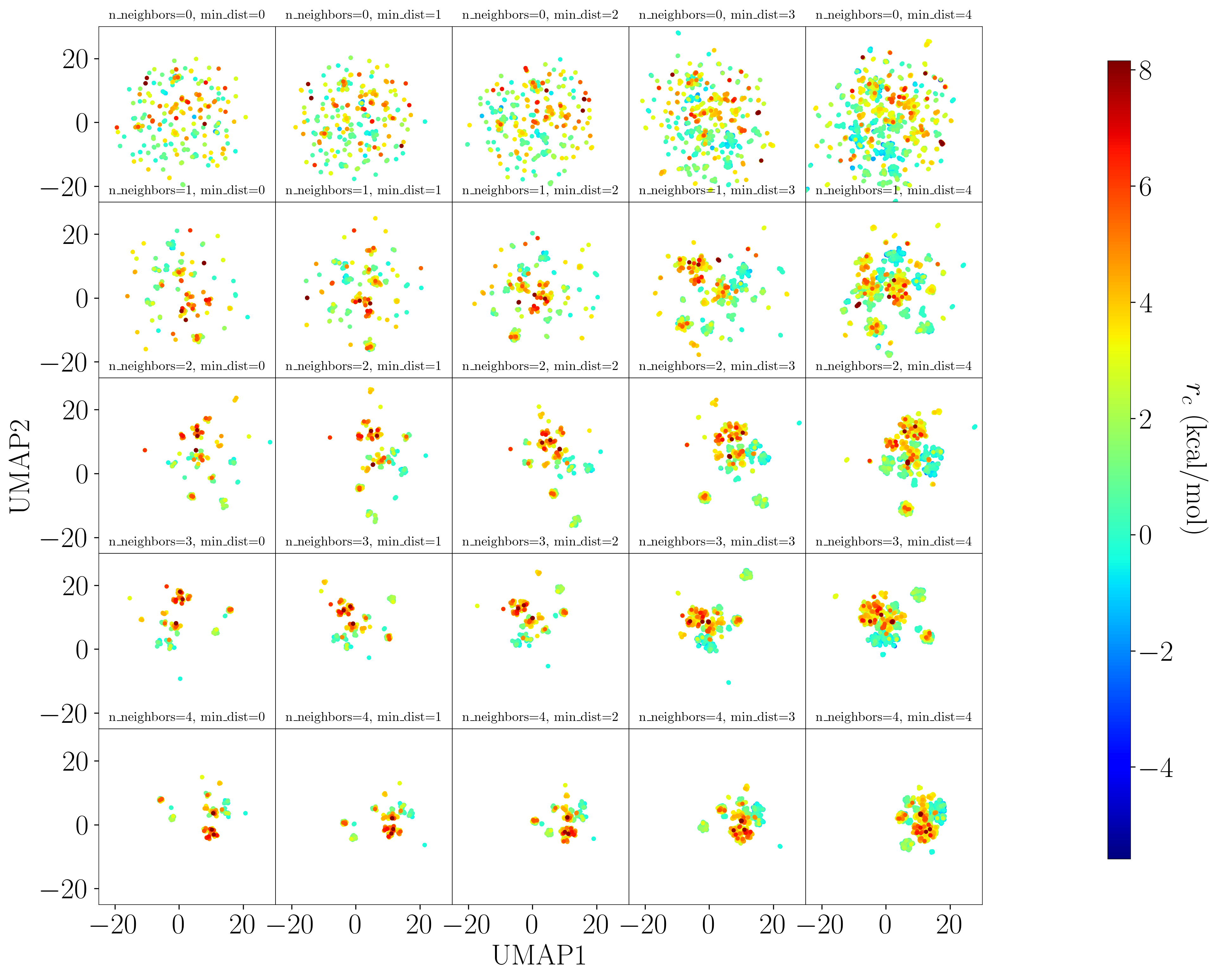}
    \caption{\textbf{UMAP analysis.}  Screening of the dependence on the number of neighbors and the minimum distance hyperparameters of the UMAP algorithm in clustering the reference \textquoteleft E/Z dataset' based on Morgan fingerprint of Z molecules. The clusters are colored according to the energy gaps computed on DFT-optimized geometries.}
    \label{fig:UMAP_clustering} 
\end{figure}

Similarly to t-SNE, UMAP is a promising dimensions reduction technique for datasets with high-dimension data. Since the outcome of UMAP is highly affected by the choice of hyperparameters, the optimization of the two principal hyperparameters, i.e. the number of neighbors and the minimum distance between data points, was carried out. As shown in Fig. \ref{fig:UMAP_clustering}, the shapes of the clusters dramatically change according to the choice of different hyperparameters. Moreover, in none of the dimension reductions reported in Fig. \ref{fig:UMAP_clustering} a clear improvement in the description of the structure-property relationship is reached.

To summarize, PCA provides a robust and fast transformation of data. t-SNE is a good clustering method, but it does not retrieve any information extra to PCA, whereas increases the computational time. UMAP does not improve either the interpretation of the actual distribution of molecules in different clusters and the choice of hyperparameters is far to be trivial.

\FloatBarrier
\subsection{Simulations Details}
In the following, details regarding the time-evolution of the density of valid solutions, entropy of each model, and hyperparameters used in each PROTEUS simulation are reported.

\subsubsection{Hyperparameters}
Tab. \ref{tab:hyperparameters} reports the hyperparameters used for all the simulations discussed in the present work.

\begin{table}[!htbp] 
\centering
\caption{Full list of hyperparameters used in each PROTEUS experiment.}\label{tab:hyperparameters}%
\begin{tabular}{@{}cccccccc@{}}
\toprule
\textbf{Name} & \textbf{Symbol} & \multicolumn{6}{c}{\textbf{Value}} \\ [0.5ex]
\midrule
Backbone-E & - & \multicolumn{6}{c}{\texttt{Ca1C(CCCCa1)CEC}} \\
Backbone-Z & - & \multicolumn{6}{c}{\texttt{Ca1C(CCCCa1)CZC}} \\
Vocabulary single & $V_S$ & \multicolumn{6}{c}{$[$\texttt{Z}, \texttt{E}, \texttt{\#}, \texttt{C}, \texttt{F}, \texttt{N}, \texttt{O}$]$} \\
Vocabulary double & $V_D$ & \multicolumn{6}{c}{$[$\texttt{()}, \texttt{11}, \texttt{a1a1}$]$} \\
Tokens & $L$ & 4 & 5 & 6 & 6 & 6 & 7 \\
Seed & - & [1, 2, 3] & [1, 2, 3] & [1, 2, 5] & 1 & 1 & 1 \\
Batch size & $B$ & 16 & 16 & 16 & 16 & 16 & 16 \\
Learning rate & $\eta$ & 1.e-5 & 1.e-5 & 1.e-5 & 1.e-5 & 1.e-5 & 1.e-5 \\
Discount & $\gamma$ & 1 & 1 & 1 & 1 & 1 & 1 \\
\makecell[c]{PPO clip\\coefficient} & $\epsilon$ & 0.2 & 0.2 & 0.2 & 0.2 & 0.2 & 0.2 \\
\makecell[c]{Master entropy\\coefficient} & ${c_e}_{M}$ & 5.e-3 & 5.e-3 & 5.e-3 & 5.e-3 & 6.e-3 & 6.e-3 \\
\makecell[c]{Position models\\entropy coefficient} & ${c_e}_{P}$ & 1.e-3 & 1.e-3 & 1.e-3 & 1.e-3 & 1.e-3 & 1.e-3 \\
\makecell[c]{Single generator\\entropy coefficient} & ${c_e}_{G^S}$ & 4.e-2 & 4.e-2 & 4.e-2 & 4.e-2 & 4.e-2 & 4.e-2 \\
\makecell[c]{Double generator\\entropy coefficient} & ${c_e}_{G^D}$ & 1.e-2 & 1.e-2 & 1.e-2 & 1.e-2 & 1.e-2 & 1.e-2 \\
\makecell[c]{SMILES memory\\length} & $n$ & 15 & 15 & 15 & 15 & 15 & 15 \\
\makecell[c]{Top SMILES\\memory length} & $K$ & 3 & 3 & 3 & 3 & 3 & 3 \\
Coefficient $r_c$ & $\alpha$ & 1 & 1 & 1 & 1 & 1 & 1 \\
Coefficient $r_d$ & $\beta$ & 1 & 1 & 1 & 0 & 1 & 1 \\
Coefficient $r_d$ & $\beta$ & 1 & 1 & 1 & 1 & 1 & 1 \\
Coefficient $r_d$ & $\beta$ & 1 & 1 & 1 & 2 & 1 & 1 \\
Energy gap & $\Delta E$ & E/Z & E/Z & E/Z & \textit{trans}/\textit{cis} & \textit{cis}/\textit{trans} & E/Z \\
\bottomrule
\end{tabular}
\end{table}

\FloatBarrier
\clearpage
\subsubsection{The 4-token simulations}
\begin{figure}[!htbp] 
    \centering
    \includegraphics[width=1.\linewidth]{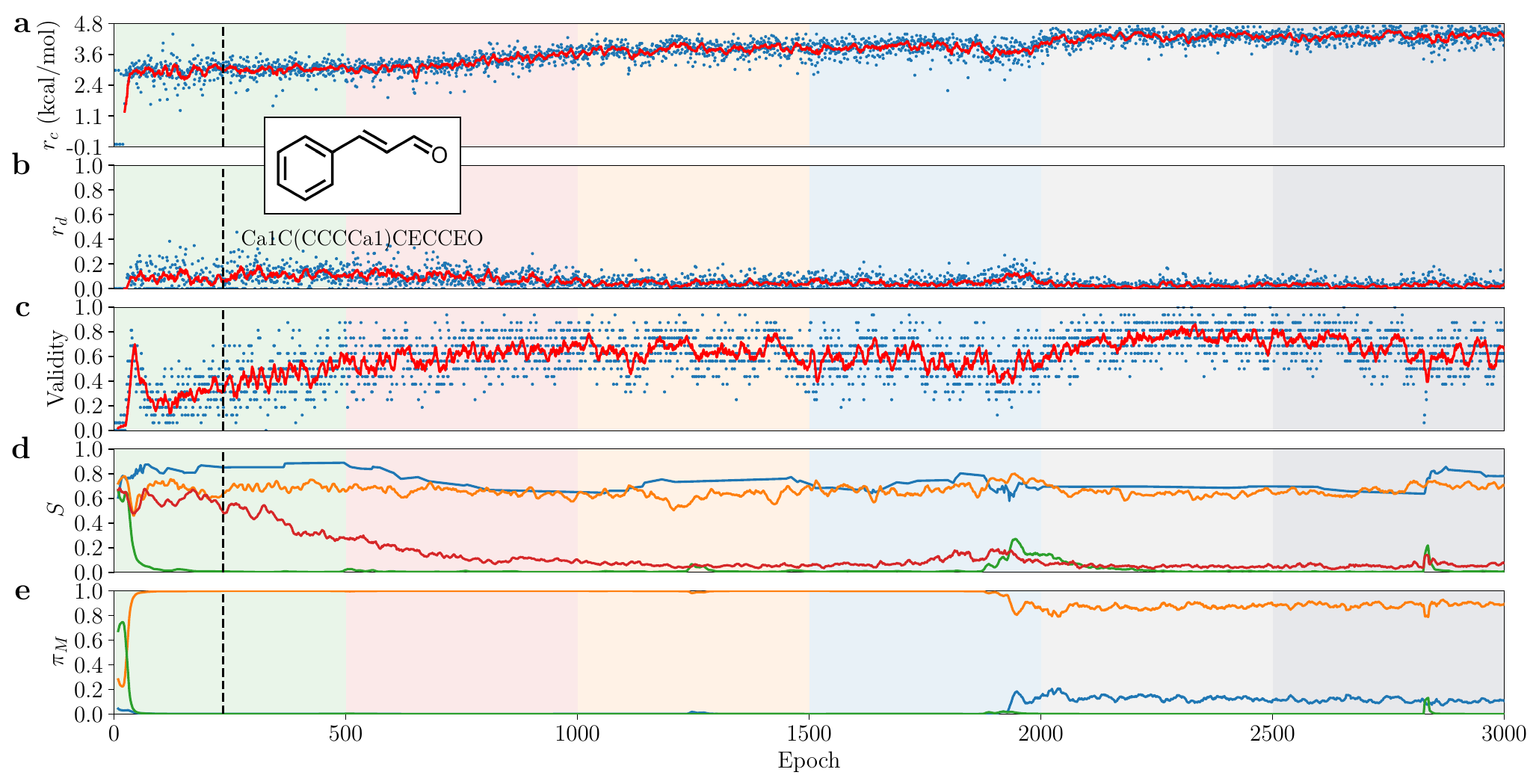}
    \caption{\textbf{Simulation 1.} Time-evolution of (\textbf{a}) the chemical and (\textbf{b}) the diversity reward, and (\textbf{c}) the validity density. Both the mean value of each epoch (blue scatter) and the running average (solid red line) are reported. Time-evolving average of the (\textbf{d}) entropy value of each policy of the models - $M$ (blue), $G^S$ (yellow), $G^D$ (green), $P$ (red) - and the (\textbf{e}) policy value of the master $M$, \texttt{add double-char} (blue), \texttt{add single-char} (yellow), \texttt{return state} (green). The epoch corresponding to the first generation of the best molecule (inset), as ranked in the \textquoteleft E/Z dataset', is marked with a dashed line. The total epochs shown in each panel are divided into 6 subsets, as highlighted by different color backgrounds.}
    \label{fig:entropies_sim1} 
\end{figure}

\begin{figure}[!htbp] 
    \centering
    \includegraphics[width=1.\linewidth]{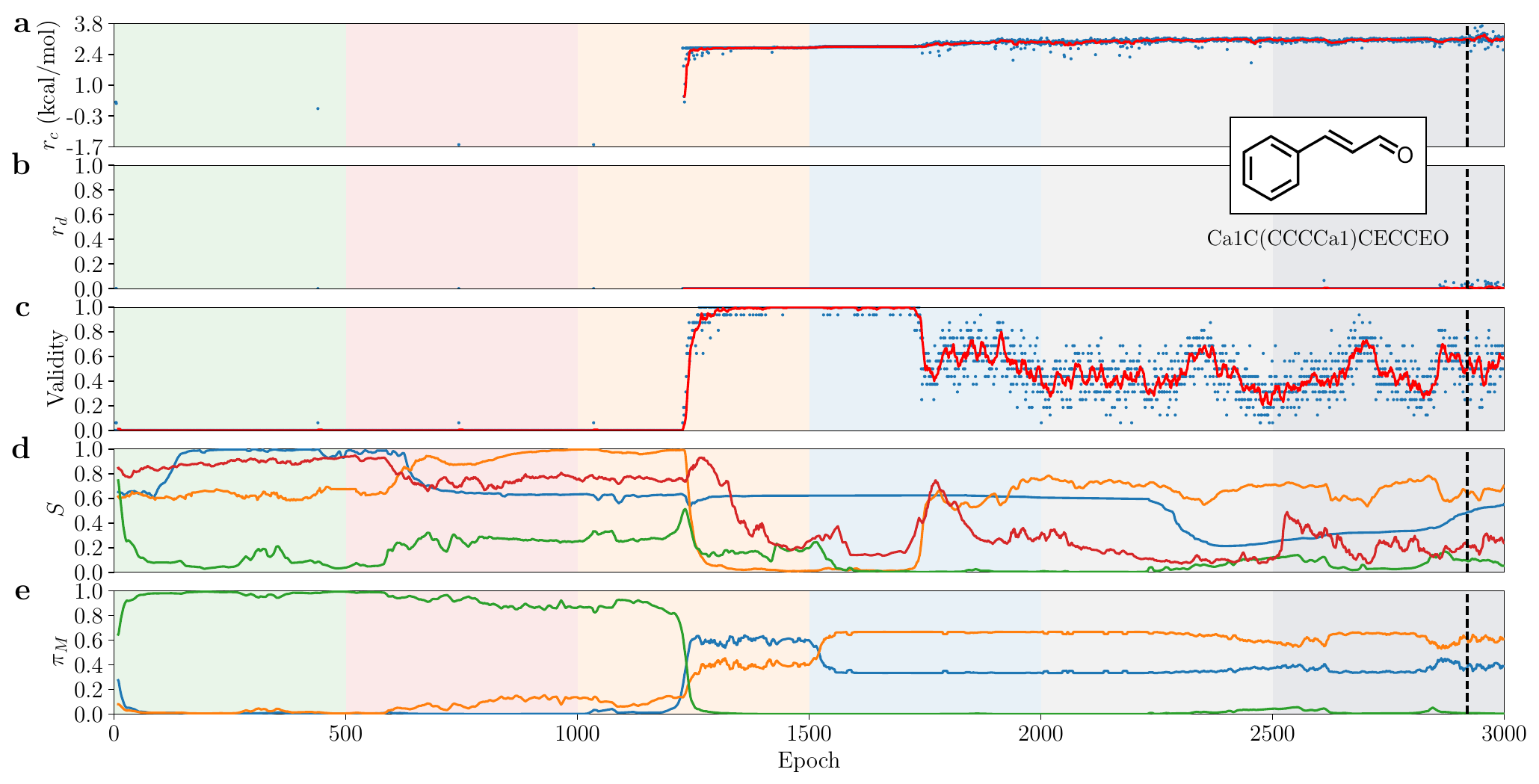}   
    \caption{\textbf{Simulation 2.} Time-evolution of (\textbf{a}) the chemical and (\textbf{b}) the diversity reward, and (\textbf{c}) the validity density. Both the mean value of each epoch (blue scatter) and the running average (solid red line) are reported. Time-evolving average of the (\textbf{d}) entropy value of each policy of the models - $M$ (blue), $G^S$ (yellow), $G^D$ (green), $P$ (red) - and the (\textbf{e}) policy value of the master $M$, \texttt{add double-char} (blue), \texttt{add single-char} (yellow), \texttt{return state} (green). The epoch corresponding to the first generation of the best molecule (inset), as ranked in the \textquoteleft E/Z dataset', is marked with a dashed line. The total epochs shown in each panel are divided into 6 subsets, as highlighted by different color backgrounds.}
    \label{fig:entropies_sim2} 
\end{figure}

\begin{figure}[!htbp] 
    \centering
    \includegraphics[width=1.\linewidth]{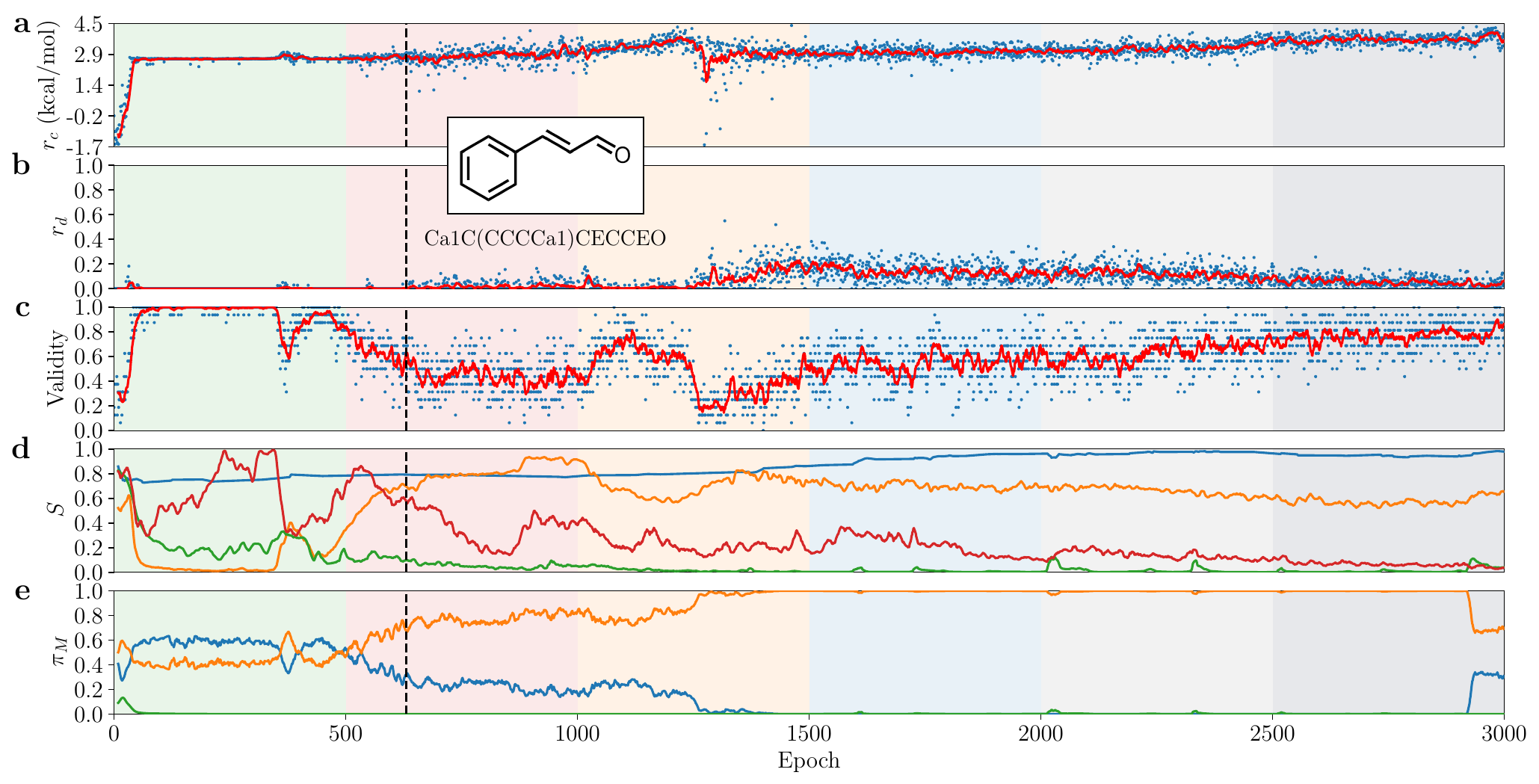}
    \caption{\textbf{Simulation 3.} Time-evolution of (\textbf{a}) the chemical and (\textbf{b}) the diversity reward, and (\textbf{c}) the validity density. Both the mean value of each epoch (blue scatter) and the running average (solid red line) are reported. Time-evolving average of the (\textbf{d}) entropy value of each policy of the models - $M$ (blue), $G^S$ (yellow), $G^D$ (green), $P$ (red) - and the (\textbf{e}) policy value of the master $M$, \texttt{add double-char} (blue), \texttt{add single-char} (yellow), \texttt{return state} (green). The epoch corresponding to the first generation of the best molecule (inset), as ranked in the \textquoteleft E/Z dataset', is marked with a dashed line. The total epochs shown in each panel are divided into 6 subsets, as highlighted by different color backgrounds.}
    \label{fig:entropies_sim3} 
\end{figure}

\FloatBarrier
\clearpage
\subsubsection{The 5-token simulations}
\begin{figure}[!htbp] 
    \includegraphics[width=1.\linewidth]{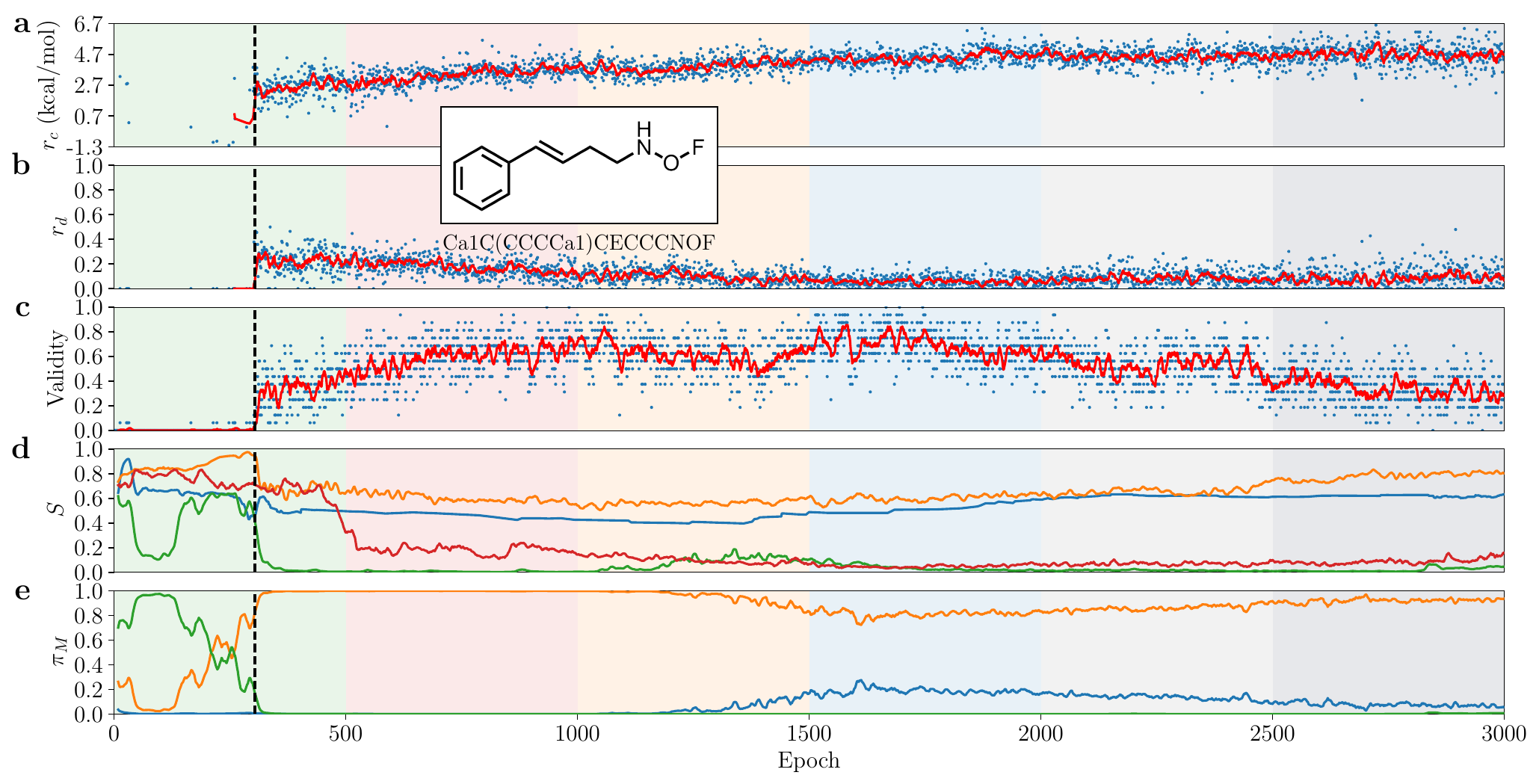}
    \caption{\textbf{Simulation 4.} Time-evolution of (\textbf{a}) the chemical and (\textbf{b}) the diversity reward, and (\textbf{c}) the validity density. Both the mean value of each epoch (blue scatter) and the running average (solid red line) are reported. Time-evolving average of the (\textbf{d}) entropy value of each policy of the models - $M$ (blue), $G^S$ (yellow), $G^D$ (green), $P$ (red) - and the (\textbf{e}) policy value of the master $M$, \texttt{add double-char} (blue), \texttt{add single-char} (yellow), \texttt{return state} (green). The epoch corresponding to the first generation of the best molecule (inset), as ranked in the \textquoteleft E/Z dataset', is marked with a dashed line. The total epochs shown in each panel are divided into 6 subsets, as highlighted by different color backgrounds.}
    \label{fig:entropies_sim4} 
\end{figure}

\begin{figure}[!htbp] 
    \centering
    \includegraphics[width=1.\linewidth]{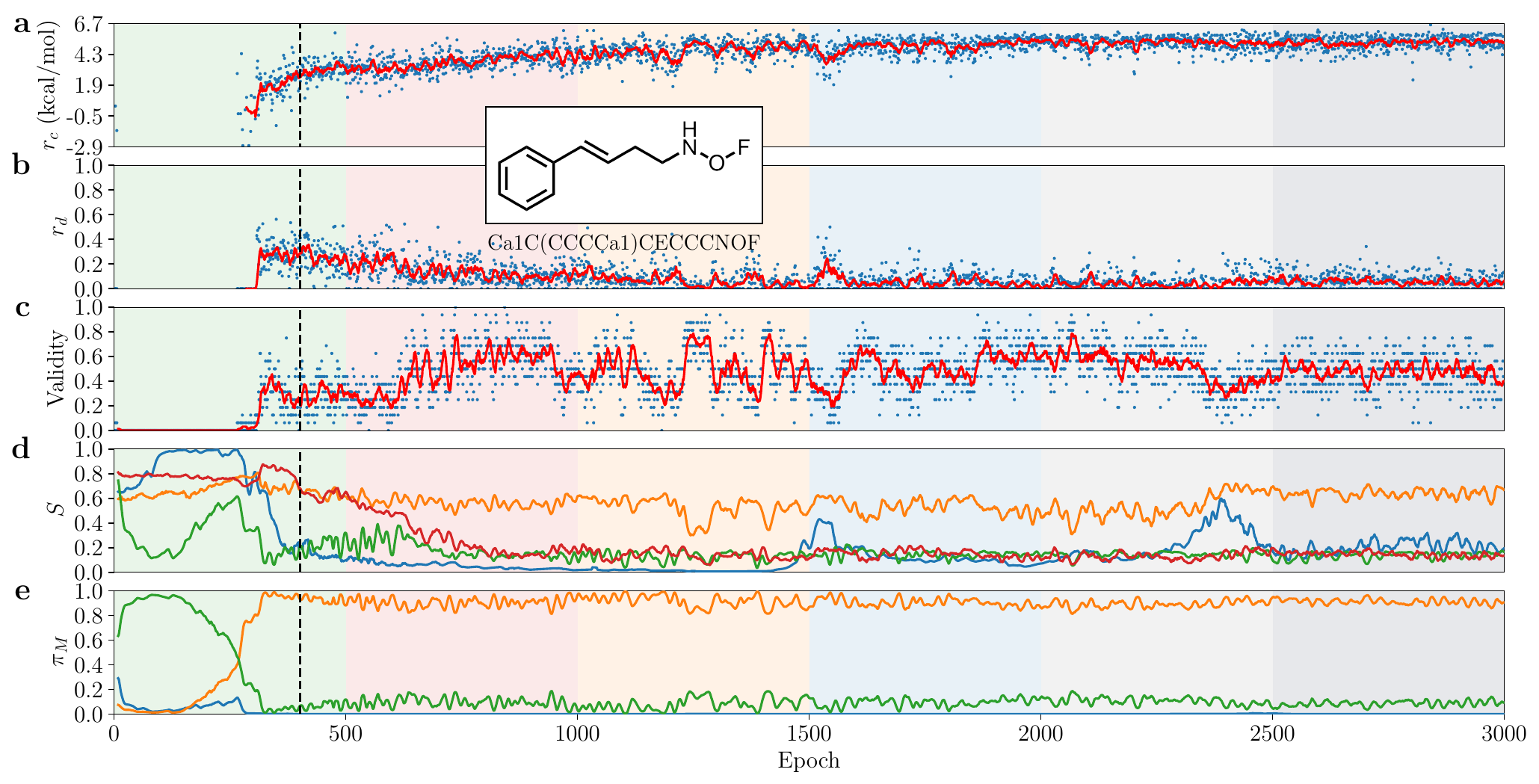}
    \caption{\textbf{Simulation 5.} Time-evolution of (\textbf{a}) the chemical and (\textbf{b}) the diversity reward, and (\textbf{c}) the validity density. Both the mean value of each epoch (blue scatter) and the running average (solid red line) are reported. Time-evolving average of the (\textbf{d}) entropy value of each policy of the models - $M$ (blue), $G^S$ (yellow), $G^D$ (green), $P$ (red) - and the (\textbf{e}) policy value of the master $M$, \texttt{add double-char} (blue), \texttt{add single-char} (yellow), \texttt{return state} (green). The epoch corresponding to the first generation of the best molecule (inset), as ranked in the \textquoteleft E/Z dataset', is marked with a dashed line. The total epochs shown in each panel are divided into 6 subsets, as highlighted by different color backgrounds.}
    \label{fig:entropies_sim5}
\end{figure}

\begin{figure}[!htbp] 
    \centering
    \includegraphics[width=1.\linewidth]{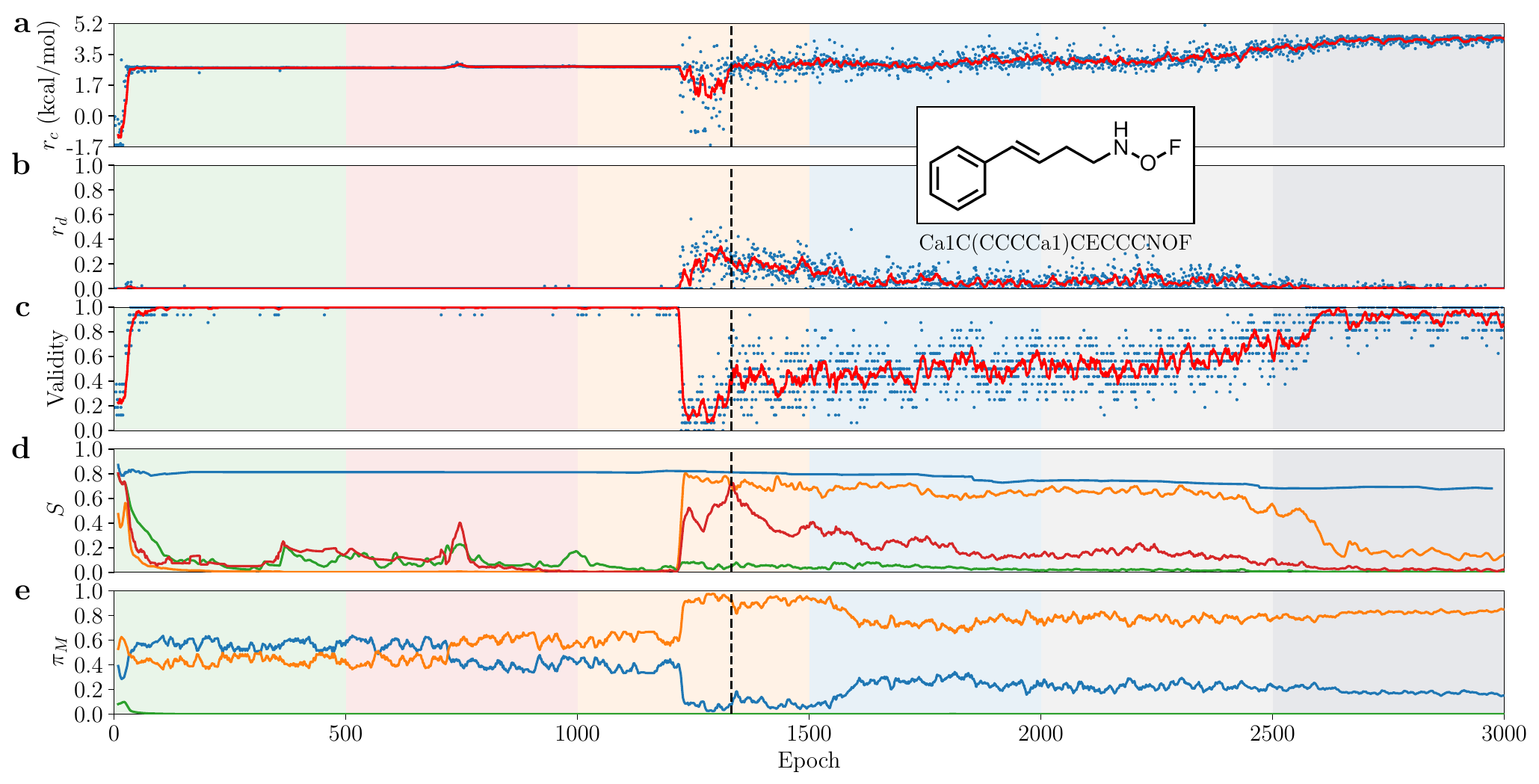}
    \caption{\textbf{Simulation 6.} Time-evolution of (\textbf{a}) the chemical and (\textbf{b}) the diversity reward, and (\textbf{c}) the validity density. Both the mean value of each epoch (blue scatter) and the running average (solid red line) are reported. Time-evolving average of the (\textbf{d}) entropy value of each policy of the models - $M$ (blue), $G^S$ (yellow), $G^D$ (green), $P$ (red) - and the (\textbf{e}) policy value of the master $M$, \texttt{add double-char} (blue), \texttt{add single-char} (yellow), \texttt{return state} (green). The epoch corresponding to the first generation of the best molecule (inset), as ranked in the \textquoteleft E/Z dataset', is marked with a dashed line. The total epochs shown in each panel are divided into 6 subsets, as highlighted by different color backgrounds.}
    \label{fig:entropies_sim6}
\end{figure}

\FloatBarrier
\clearpage
\subsubsection{The 6-token simulations}

\begin{figure}[!htbp] 
    \centering
    \includegraphics[width=1.\linewidth]{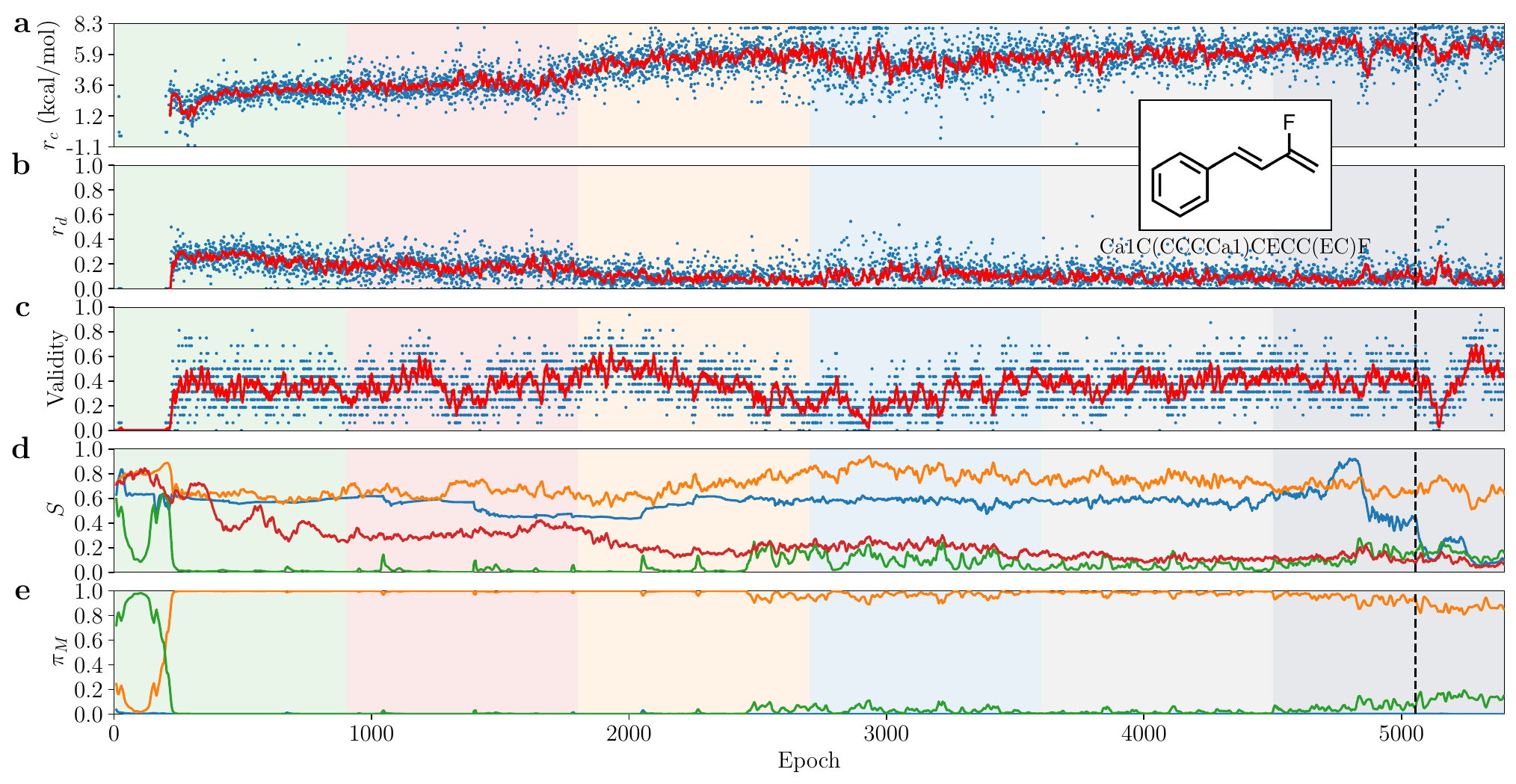}
    \caption{\textbf{Simulation 7.} Time-evolution of (\textbf{a}) the chemical and (\textbf{b}) the diversity reward, and (\textbf{c}) the validity density. Both the mean value of each epoch (blue scatter) and the running average (solid red line) are reported. Time-evolving average of the (\textbf{d}) entropy value of each policy of the models - $M$ (blue), $G^S$ (yellow), $G^D$ (green), $P$ (red) - and the (\textbf{e}) policy value of the master $M$, \texttt{add double-char} (blue), \texttt{add single-char} (yellow), \texttt{return state} (green). The epoch corresponding to the first generation of the best molecule (inset), as ranked in the \textquoteleft E/Z dataset', is marked with a dashed line. The total epochs shown in each panel are divided into 6 subsets, as highlighted by different color backgrounds.}
    \label{fig:entropies_sim7} 
\end{figure}

\begin{figure}[!htbp] 
    \centering
    \includegraphics[width=1.\linewidth]{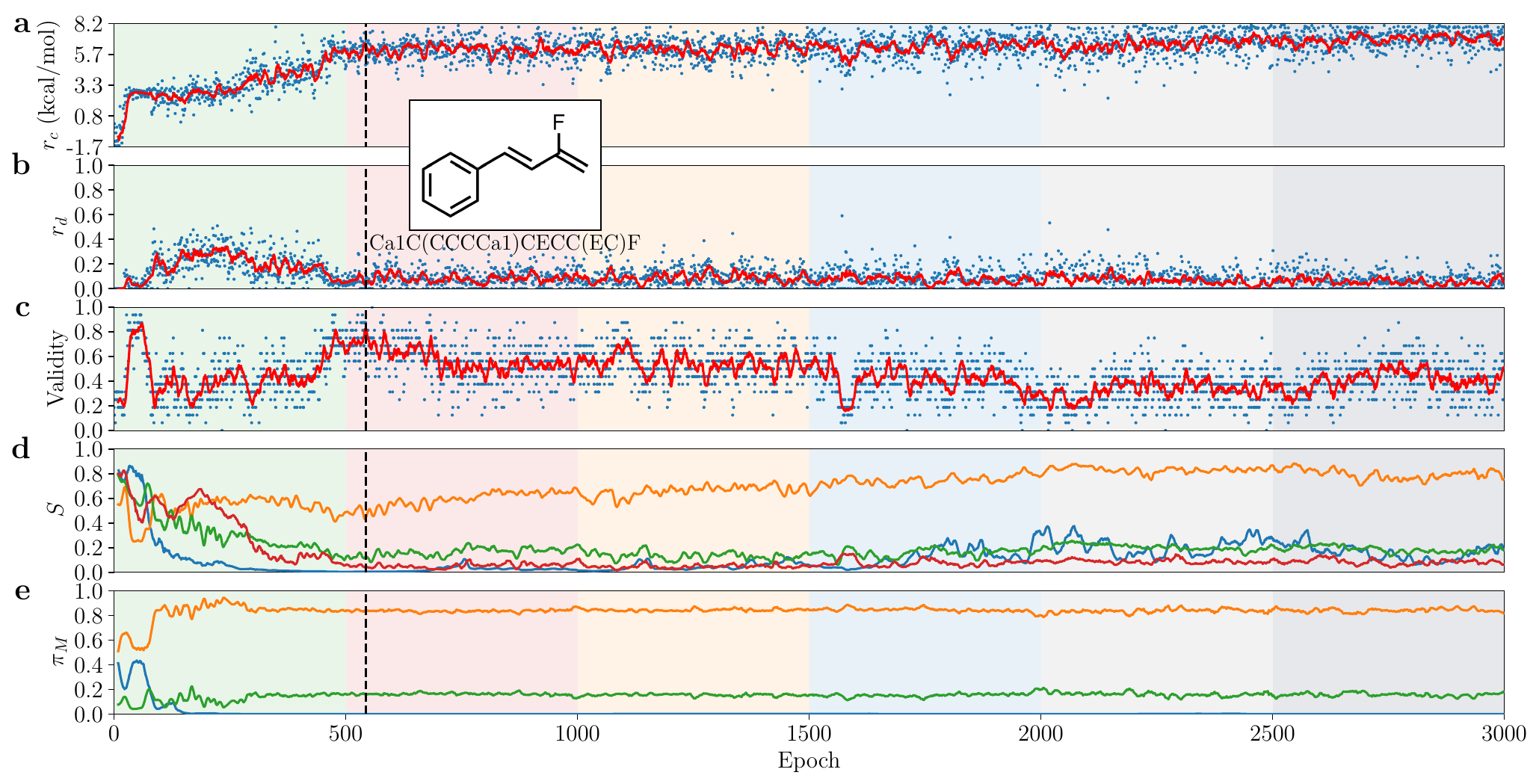}
    \caption{\textbf{Simulation 8.} Time-evolution of (\textbf{a}) the chemical and (\textbf{b}) the diversity reward, and (\textbf{c}) the validity density. Both the mean value of each epoch (blue scatter) and the running average (solid red line) are reported. Time-evolving average of the (\textbf{d}) entropy value of each policy of the models - $M$ (blue), $G^S$ (yellow), $G^D$ (green), $P$ (red) - and the (\textbf{e}) policy value of the master $M$, \texttt{add double-char} (blue), \texttt{add single-char} (yellow), \texttt{return state} (green). The epoch corresponding to the first generation of the best molecule (inset), as ranked in the \textquoteleft E/Z dataset', is marked with a dashed line. The total epochs shown in each panel are divided into 6 subsets, as highlighted by different color backgrounds.}
    \label{fig:entropies_sim8}
\end{figure}

\begin{figure}[!htbp] 
    \centering
    
    \includegraphics[width=1.\linewidth]{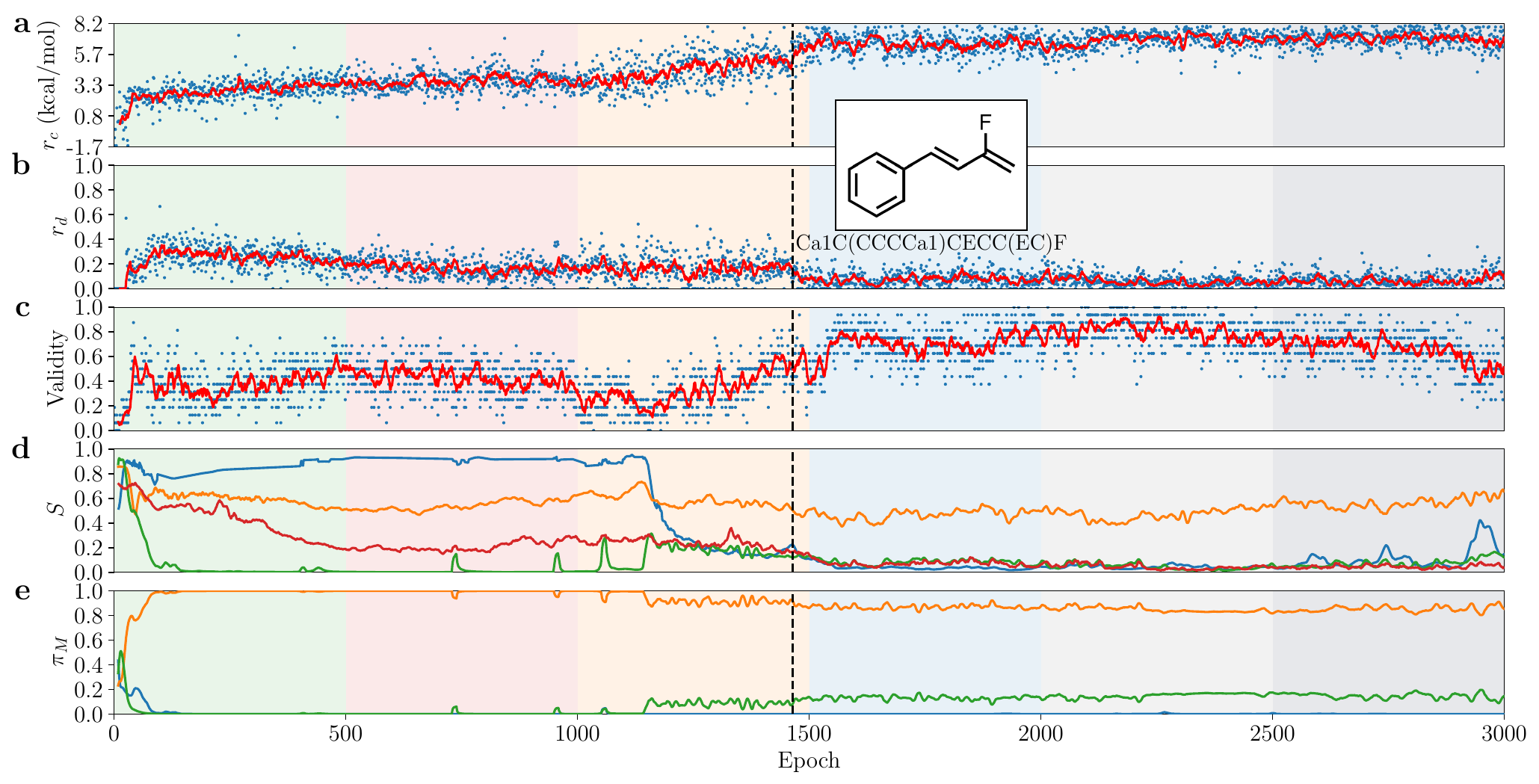}
    \caption{\textbf{Simulation 9.} Time-evolution of (\textbf{a}) the chemical and (\textbf{b}) the diversity reward, and (\textbf{c}) the validity density. Both the mean value of each epoch (blue scatter) and the running average (solid red line) are reported. Time-evolving average of the (\textbf{d}) entropy value of each policy of the models - $M$ (blue), $G^S$ (yellow), $G^D$ (green), $P$ (red) - and the (\textbf{e}) policy value of the master $M$, \texttt{add double-char} (blue), \texttt{add single-char} (yellow), \texttt{return state} (green). The epoch corresponding to the first generation of the best molecule (inset), as ranked in the \textquoteleft E/Z dataset', is marked with a dashed line. The total epochs shown in each panel are divided into 6 subsets, as highlighted by different color backgrounds.}
    \label{fig:entropies_sim9} 
\end{figure}

\FloatBarrier
\clearpage
\subsubsection{The \textit{trans}/\textit{cis} 6-token simulations}
\begin{figure}[!htbp] 
    \centering
    \includegraphics[width=1.\linewidth]{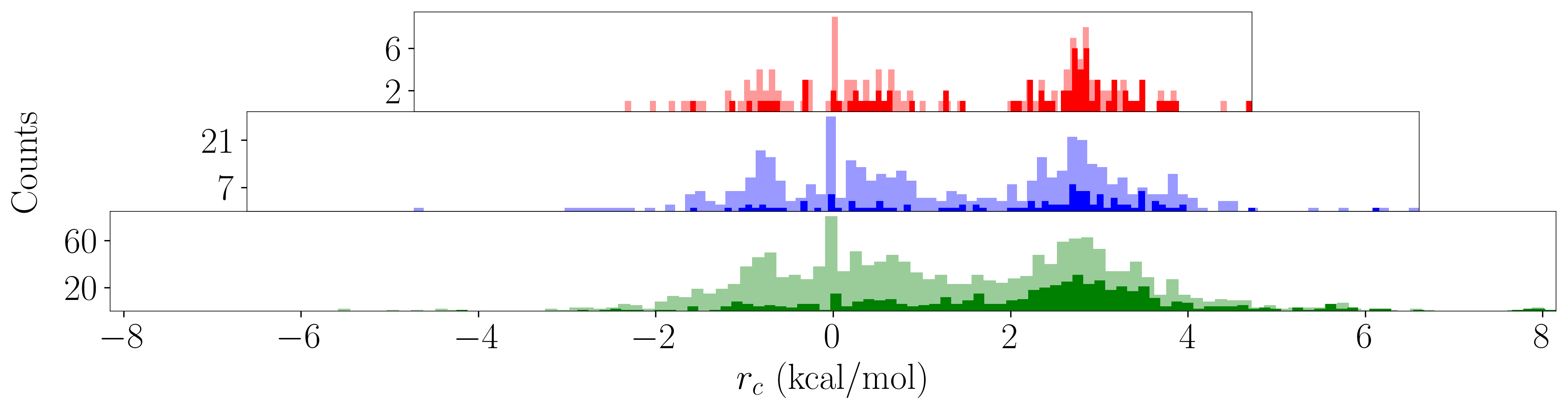}
    \caption{\textbf{Explored sub-spaces}. Explored chemical space (solid) VS reference chemical space (shadowed) for 4 (red), 5 (blue), and 6 (green) token sizes of the P-SMILES strings.}
    \label{fig:exploration_CT}
\end{figure}

\begin{figure}[!htbp] 
    \centering
    \includegraphics[width=1.\linewidth]{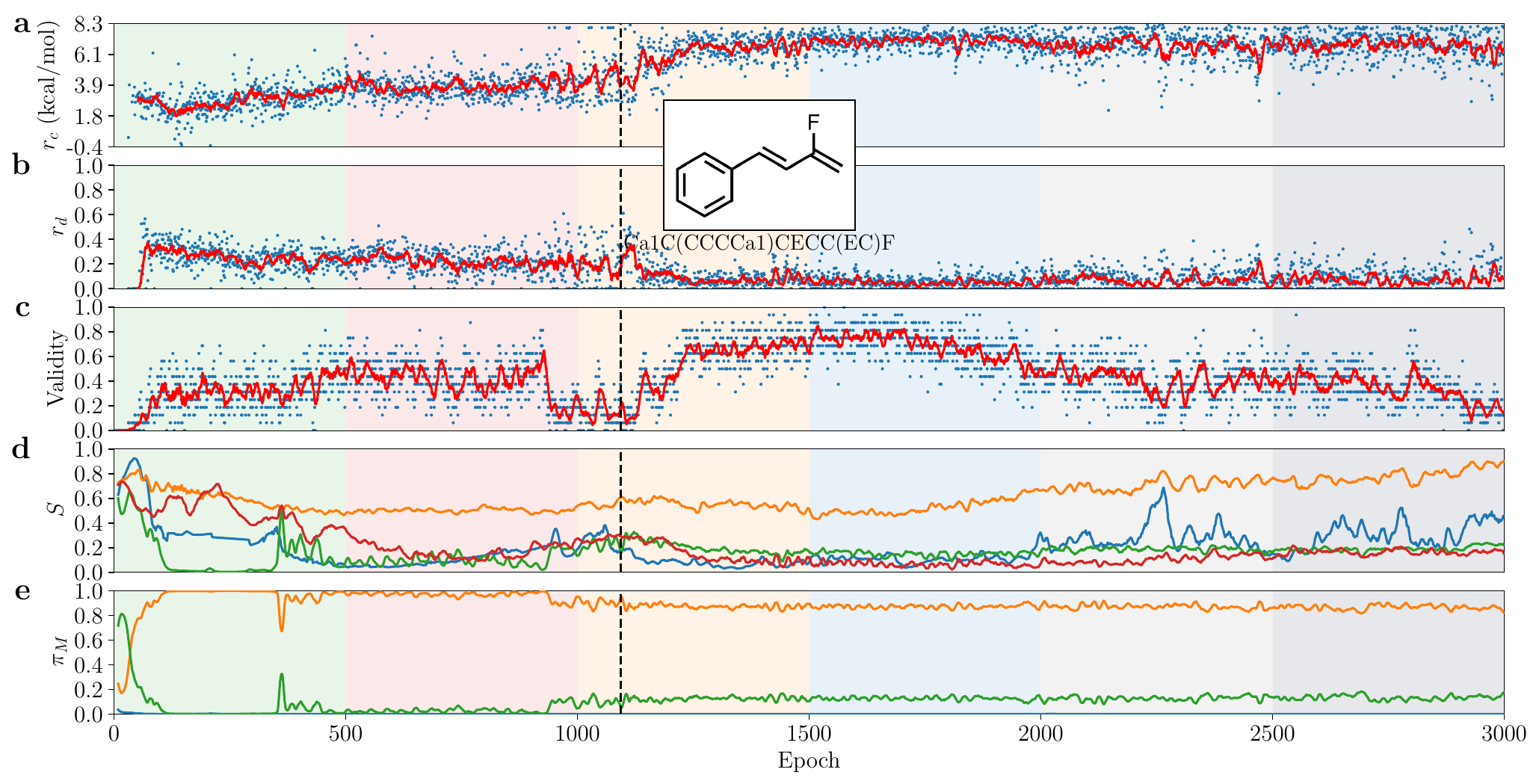}
    \caption{\textbf{Simulation 10.} Time-evolution of (\textbf{a}) the chemical and (\textbf{b}) the diversity reward, and (\textbf{c}) the validity density. Both the mean value of each epoch (blue scatter) and the running average (solid red line) are reported. Time-evolving average of the (\textbf{d}) entropy value of each policy of the models - $M$ (blue), $G^S$ (yellow), $G^D$ (green), $P$ (red) - and the (\textbf{e}) policy value of the master $M$, \texttt{add double-char} (blue), \texttt{add single-char} (yellow), \texttt{return state} (green). The epoch corresponding to the first generation of the best molecule (inset), as ranked in the \textquoteleft E/Z dataset', is marked with a dashed line. The total epochs shown in each panel are divided into 6 subsets, as highlighted by different color backgrounds.}
    \label{fig:entropies_sim10} 
\end{figure}

\begin{figure}[!htbp] 
    \centering
    \includegraphics[width=1.\linewidth]{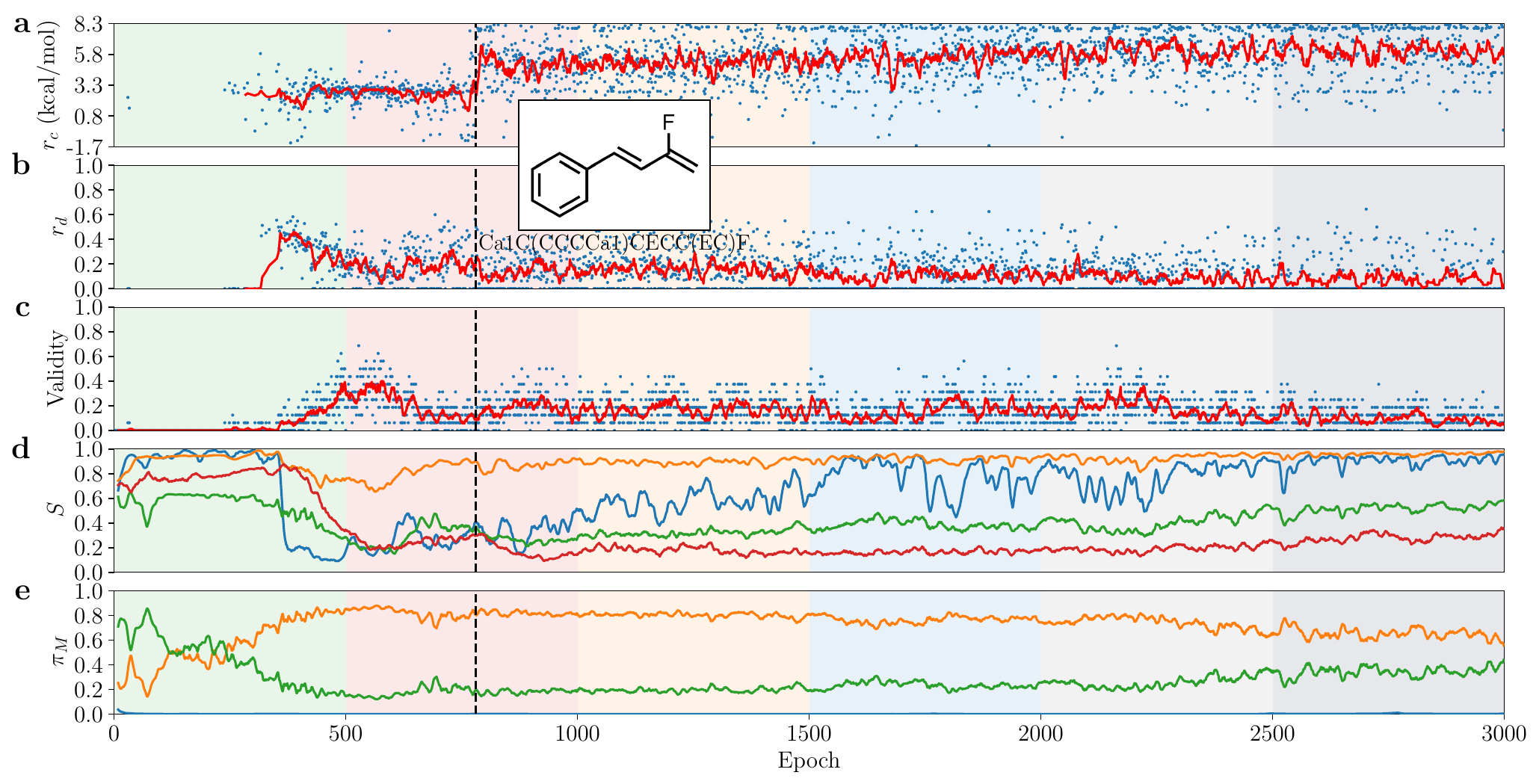}
    \caption{\textbf{Simulation 13.} Time-evolution of (\textbf{a}) the chemical and (\textbf{b}) the diversity reward, and (\textbf{c}) the validity density. Both the mean value of each epoch (blue scatter) and the running average (solid red line) are reported. Time-evolving average of the (\textbf{d}) entropy value of each policy of the models - $M$ (blue), $G^S$ (yellow), $G^D$ (green), $P$ (red) - and the (\textbf{e}) policy value of the master $M$, \texttt{add double-char} (blue), \texttt{add single-char} (yellow), \texttt{return state} (green). The epoch corresponding to the first generation of the best molecule (inset), as ranked in the \textquoteleft E/Z dataset', is marked with a dashed line. The total epochs shown in each panel are divided into 6 subsets, as highlighted by different color backgrounds.}
    \label{fig:entropies_sim13} 
\end{figure}

\begin{figure}[!htbp] 
    \centering
    \includegraphics[width=1.\linewidth]{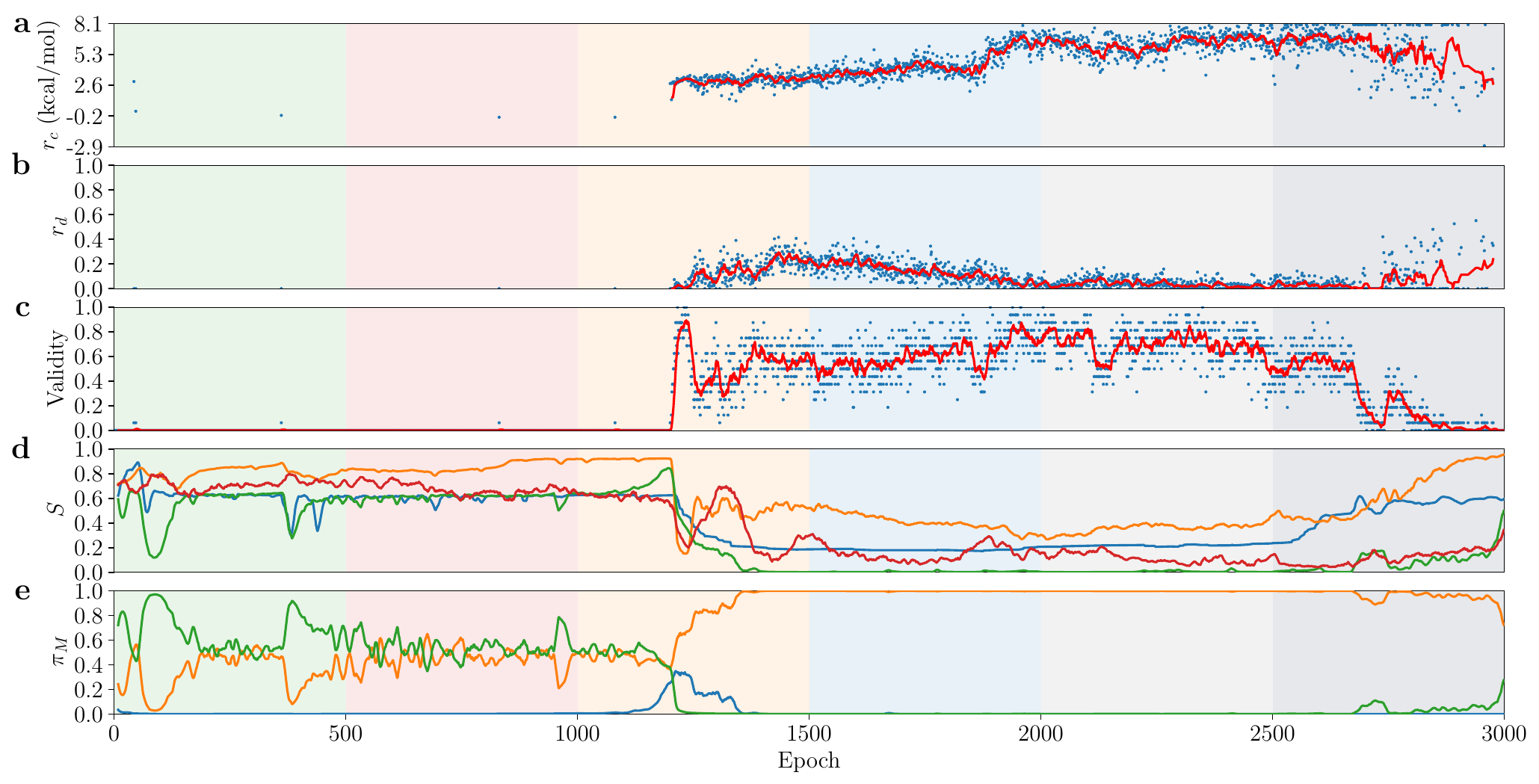}
    \caption{\textbf{Simulation 14.} Time-evolution of (\textbf{a}) the chemical and (\textbf{b}) the diversity reward, and (\textbf{c}) the validity density. Both the mean value of each epoch (blue scatter) and the running average (solid red line) are reported. Time-evolving average of the (\textbf{d}) entropy value of each policy of the models - $M$ (blue), $G^S$ (yellow), $G^D$ (green), $P$ (red) - and the (\textbf{e}) policy value of the master $M$, \texttt{add double-char} (blue), \texttt{add single-char} (yellow), \texttt{return state} (green). The epoch corresponding to the first generation of the best molecule (inset), as ranked in the \textquoteleft E/Z dataset', is marked with a dashed line. The total epochs shown in each panel are divided into 6 subsets, as highlighted by different color backgrounds.}
    \label{fig:entropies_sim14} 
\end{figure}
    
\FloatBarrier
\clearpage
\subsubsection{The \textit{trans}/\textit{cis} 7-token simulation}
Along the 7-token simulation, PROTEUS explored 5,762 syntactically valid molecules featuring up to 7 tokens. This subset is composed as follows: 
\begin{itemize}
    \item 4,276 out of 5,762 returned error.
    \item 1,976 out of 4,276 raised exception due to connectivity inconsistency at the end of the calculation of the $r_c$.
    \item 2,185 out of 4,276 raised exception since they are open-shell systems.
    \item 106 out of 4,276 raised exception during the xTB optimization.
    \item 9 out of 4,276 raised exception during the DFT optimization.
\end{itemize}

\begin{figure}[!htbp] 
    \centering
    \includegraphics[width=1.\linewidth]{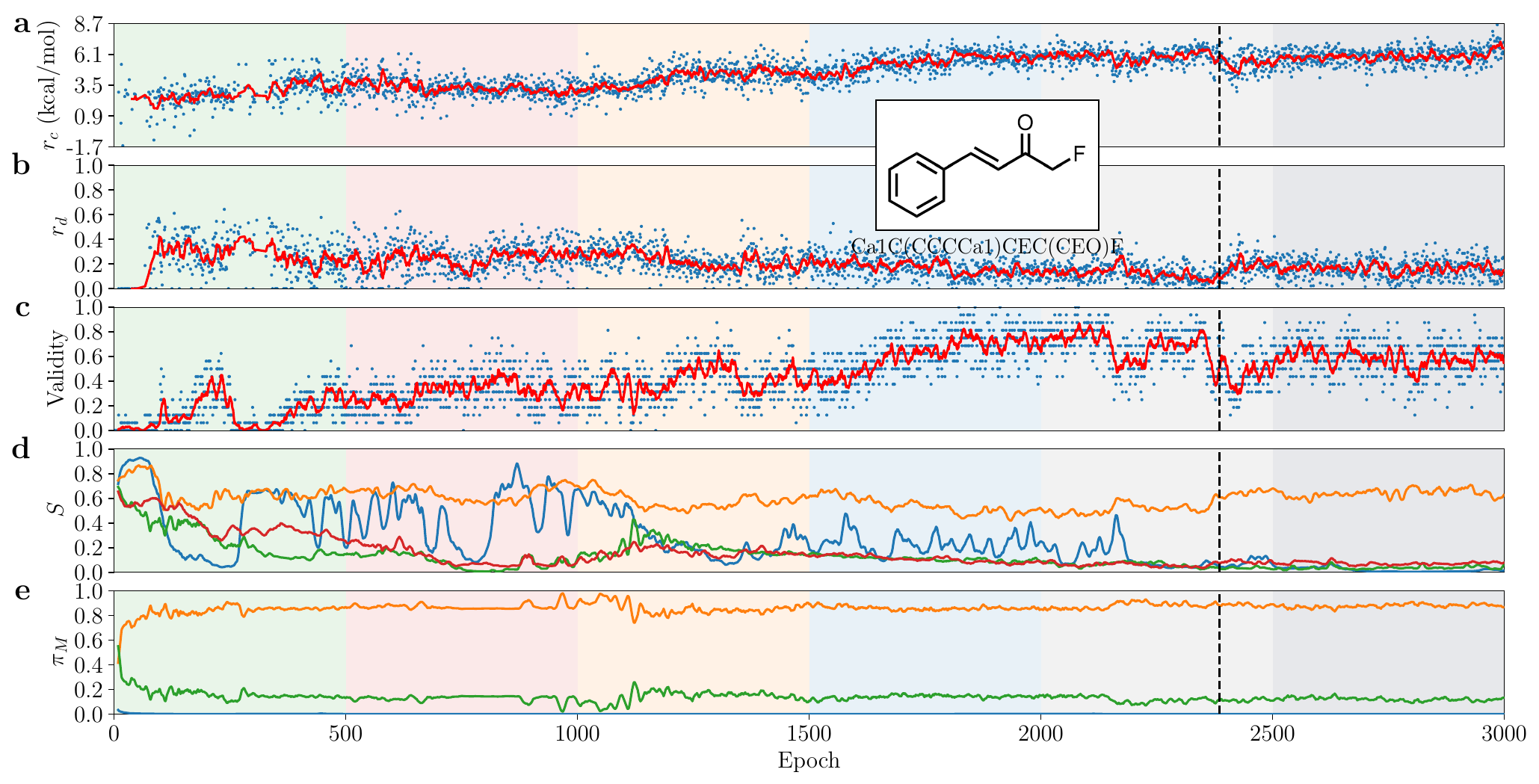}     
    \caption{\textbf{Simulation 12.} Time-evolution of (\textbf{a}) the chemical and (\textbf{b}) the diversity reward, and (\textbf{c}) the validity density. Both the mean value of each epoch (blue scatter) and the running average (solid red line) are reported. Time-evolving average of the (\textbf{d}) entropy value of each policy of the models - $M$ (blue), $G^S$ (yellow), $G^D$ (green), $P$ (red) - and the (\textbf{e}) policy value of the master $M$, \texttt{add double-char} (blue), \texttt{add single-char} (yellow), \texttt{return state} (green). The epoch corresponding to the first generation of the best molecule (inset), as ranked in the \textquoteleft E/Z dataset', is marked with a dashed line. The total epochs shown in each panel are divided into 6 subsets, as highlighted by different color backgrounds.}
    \label{fig:entropies_sim12}
\end{figure}

\FloatBarrier
\clearpage
\subsubsection{Details on the 6-token \textit{cis}/\textit{trans} simulation} \label{sec:TC_prob}

Inverse designing tailored substituents of the styrene's backbone to maximize the \textit{cis}/\textit{trans} energy gap using 6 P-SMILES tokens is highly challenging. Similarly to the \textit{trans}/\textit{cis} problem, the molecule with the largest energy gap, i.e. \texttt{Ca1C(CCCCa1)CECCONZF}, is in a low-frequency region of rewards. Moreover, as disclosed by the PCA simulation, \texttt{Ca1C(CCCCa1)CECCONZCF} is chemically and structurally more similar to molecules with much smaller energy gaps. In fact, \texttt{Ca1C(CCCCa1)CECCONZCF} does not belong to the cluster containing the molecules with positive energy differences, but to one where most of the systems feature a more stable \textit{trans} conformer than for the \textit{cis} one (see the Main text). Therefore, PROTEUS must balance exploration and exploitation to successfully solve the problem.

\begin{figure}[!htbp] 
    \centering
    \includegraphics[width=1.\linewidth]{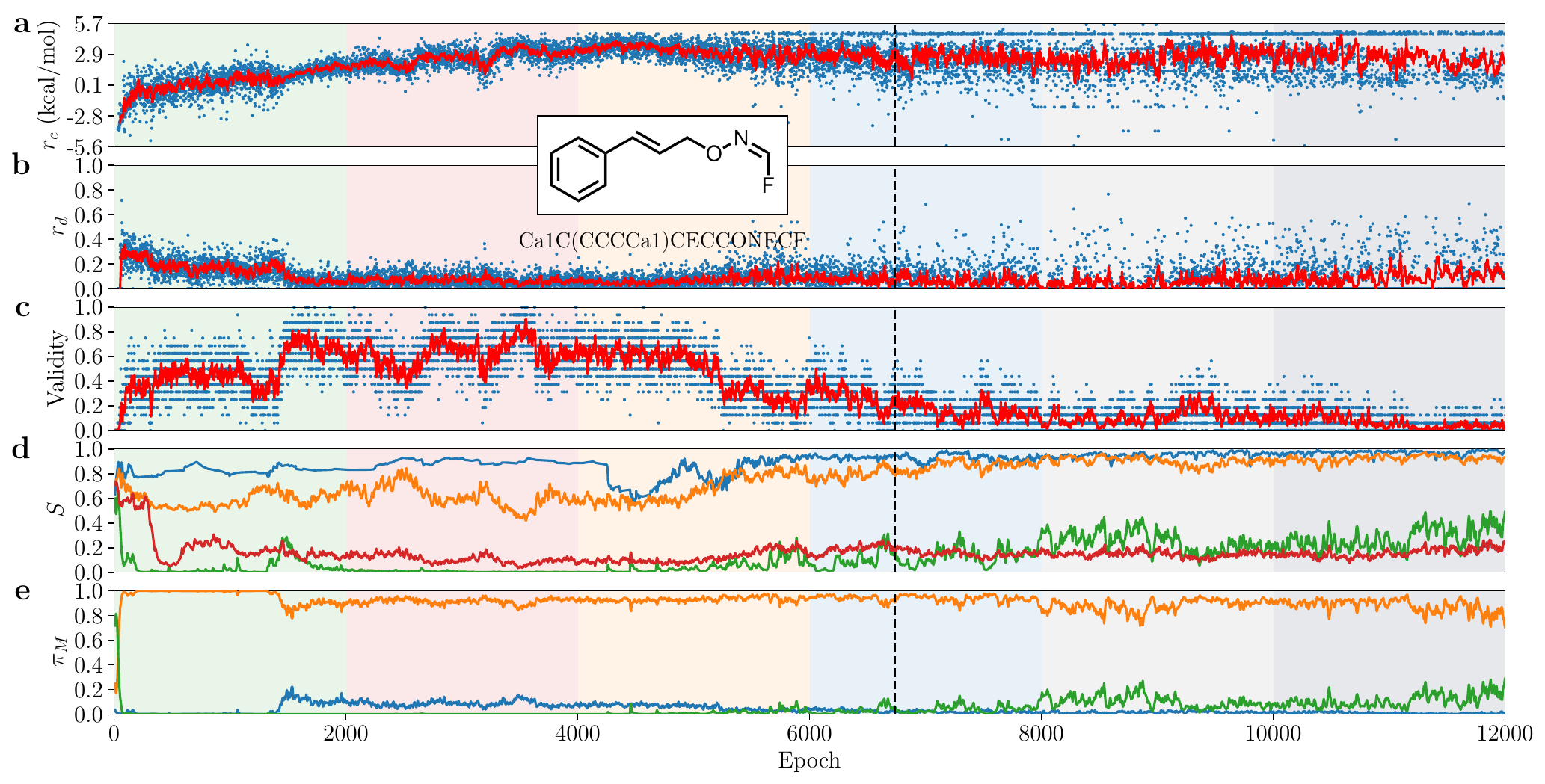}
    \caption{\textbf{Simulation 11.} Time-evolution of the \textbf{a} chemical and the \textbf{b} diversity reward, and the \textbf{c} validity density. Both the mean value of each epoch (blue scatter) and the running average (solid red line) are reported. \textbf{d} Time-evolving average of the entropy value of each policy of the models: $M$ (blue), $G^S$ (yellow), $G^D$ (green), $P$ (red). \textbf{e} Time evolving average policy value of the master $M$, \texttt{add double-char} (blue), \texttt{add single-char} (yellow), \texttt{return state} (green). The epoch corresponding to the first generation of the best molecule (inset), as ranked in the \textquoteleft E/Z dataset', is marked with a dashed line. The total epochs shown in each panel are divided into 6 subsets, as highlighted by different color backgrounds.}
    \label{fig:entropies_sim11} 
\end{figure}

Fig. \ref{fig:entropies_sim11} summarizes the results of the simulation. The overall simulation was 12,000 epochs long, instead of 3,000 as for the other simulations. The reason why we decided to use a longer simulation time is to allow PROTEUS to increase its exploration and verify if it benefits from a longer exploration phase.
During the first 4,000 epochs, PROTEUS explores the space of solutions while maximizing $r_c$. Similarly to what already discussed in Secc. \ref{sec:EZ_prob} and \ref{sec:CT_prob}, $r_c$ and $r_d$ show opposite trends: the wider the exploitation of the chemical property, the lower the diversity reward (Figg. \ref{fig:entropies_sim11}a and \ref{fig:entropies_sim11}b).
After the ca. 5,000 epochs, PROTEUS prioritizes back the exploration of the chemical space rather than the exploitation of $r_c$. Such behavior is encouraged by an increment in the total entropy, $S$, of the master, which pushes PROTEUS to explore new regions of the space of solutions (Fig. \ref{fig:entropies_sim11}d).
In fact, the $r_d$ increases (Fig. \ref{fig:entropies_sim11}b) and the unexplored molecules are generated.
Thanks to this second exploration phase, PROTEUS successfully generates \texttt{Ca1C(CCCCa1)CECCONZCF} during the 6,734-th epoch. This result is really remarkable, since it proves again that the architecture of PROTEUS promotes the exploration of challenging topologies of the chemical space.

\FloatBarrier
\clearpage
\subsubsection{Sensitivity analysis}\label{sec:sensitivity_analysis}

The choice of the optimal values of the hyperparameters is a key ingredient in RL simulations, even if it must be highlighted that this framework is less influenced by the choice of hyperparameters than other generative techniques \cite{anstine2023}.

\begin{figure}[!htbp] 
  \centering
  \includegraphics[width=1.\linewidth]{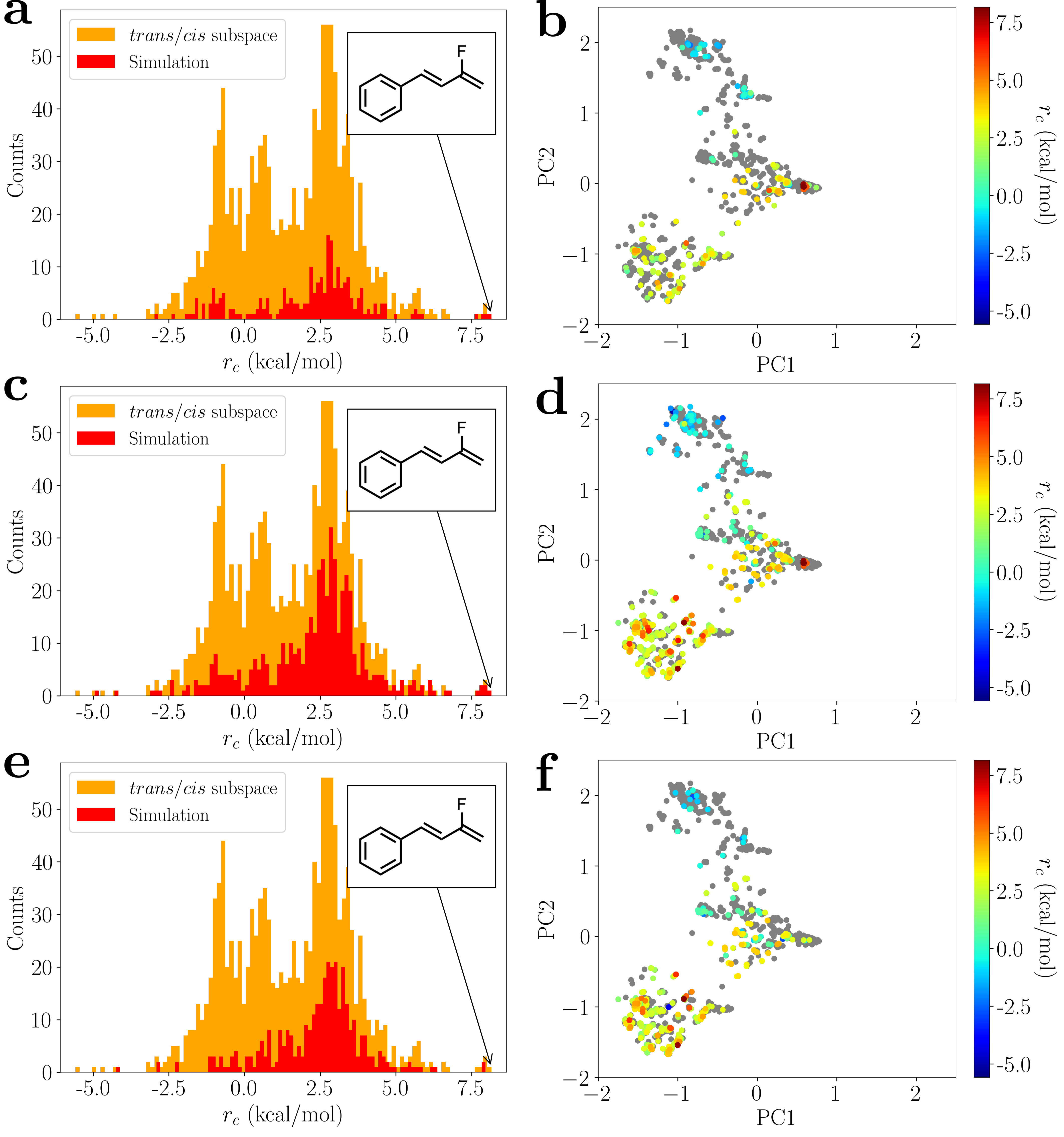}
  
  \caption{\textbf{Sensitivity analysis.} PCA analysis and rewards distributions of the exploration of the chemical space for the \textit{trans}/\textit{cis} problem according to different $\alpha:\beta$ ratios, i.e., 1:0 (\textbf{a} and \textbf{b}), 1:1 (\textbf{c} and \textbf{d}), and 1:2 (\textbf{e} and \textbf{f}).}
  \label{fig:panel7}
\end{figure}

Among others, $\alpha$ and $\beta$ govern the exploitation and the exploration of the learning process, respectively. They do this by weighting the $r_c$ and the $r_d$ values in the formulation of the reward $r_t$ (Equation \ref{eq:fitness}). Since the importance of the role played by $\alpha$ and $\beta$, we verified how different ratios of those hyperparameters influence the simulations of the \textit{trans}/\textit{cis} problem. All the other hyperparameters listed in Tab. \ref{tab:hyperparameters} are kept constant. Results are given in Tab. \ref{tab:sensitivity_analysis} and Fig. \ref{fig:panel7}.

When $\alpha : \beta = 1 : 1$, PROTEUS explores the space of solutions broadly while exploiting the task to find the best state (Fig. \ref{fig:panel7}c). During the simulation 43\% of the generated states are valid, with 464 out of the total being unique states (Tab. \ref{tab:sensitivity_analysis}).

Gratifyingly, when $\beta = 0$, i.e. the $r_d$ contribution is ignored, PROTEUS solves the \textit{trans}/\textit{cis} problem as well, but exploring a smaller part of the reference chemical space after the same number of epochs as for the previous simulation. The number of unique valid states generated is 228, while the number of total valid generated states, $\rho_v$, is 14\% of the total states (Tab. \ref{tab:sensitivity_analysis}). The lower exploration is witnessed also by the topological analysis of the explored regions (Fig. \ref{fig:panel7}a) and of the distribution of the reward (Fig. \ref{fig:panel7}b). The comparison of Fig. \ref{fig:panel7}b and Fig. \ref{fig:panel7}d highlights that when the final reward $r_t$ is corrected by the contribution of the chemical diversity, PROTEUS focuses on the broad exploration of the regions related to large reward values. 

When $\beta = 2\alpha$ (Figg. \ref{fig:panel7}e and \ref{fig:panel7}f), PROTEUS performs worse than when $\alpha : \beta = 1 : 1$. In fact, the number of unique valid states generated during the simulation drops from 464 to 338 (Tab. \ref{tab:sensitivity_analysis}) and PROTEUS does not generate the best state. The fact that the density of valid solutions lowers is ascribed to an excessive exploration of the chemical space. In fact, the $r_d$ value embeds the structural and chemical diversity between the latest generated states and the previous ones (Sec. \ref{sec:diversity}). When the reward is doped to explore the most different states possible, PROTEUS tends to generate less meaningful, i.e. valid, P-SMILES string.

To summarize, these outcomes show how the choice of a proper $\alpha : \beta$ ratio leads to different explorations of the chemical space. Moreover, these outcomes highlight that weighting the $r_t$ with $\beta r_d$ is a key critical component when exploring chemical spaces with challenging topologies.

\begin{table}[!htbp] 
\caption{\textbf{Sensitivity analysis.} Influence of the ratio of the hyperparameters $\alpha$ and $\beta$ on the \textit{trans}/\textit{cis} simulation. The density of valid states, $\rho_v$, the density of unique valid states generated, $\rho_{vu}$, and the number of unique valid states are reported.}
\label{tab:sensitivity_analysis}
\centering
\begin{tabular}{@{}cccc@{}}
\toprule
$\mathbf{\alpha : \beta}$ & \textbf{1 : 0} & \textbf{1 : 1} & \textbf{1 : 2} \\ 
\midrule
$\rho_v (\%)$             & 14            & 43            & 31            \\ 
$\rho_{vu} (\%)$          & 0.47          & 0.96          & 0.70          \\ 
Unique states             & 228           & 464           & 338           \\ 
\bottomrule
\end{tabular}
\end{table}

\newpage
\printbibliography[heading=bibintoc, title={References}]
\end{appendices}

\end{document}